%% file: main.tex
\documentclass[journal]{IEEEtran}
\input{Section_0/packages}
\begin{document}
\input{Section_0/title}
\maketitle
\input{Section_0/abstract_and_keywords}
\section{Introduction}\label{sec1}
\input{Section_I/intro}
\section{Background and Contributions}\label{sec2}
\input{Section_II/intro}
\subsection{Need for FRPs}\label{sec2a}
\input{Section_II/a}
\subsection{Review of Existing Approaches}\label{sec2b}
\input{Section_II/b}
\subsection{Research and Practical Gaps \& Our Contributions}\label{sec2c}
\input{Section_II/c}
\section{Proposed Methodology}\label{sec3}
\input{Section_III/intro}
\subsection{First Market Pass: Stochastic Unit Commitment (SUC) Problem}\label{sec3a}
\input{Section_III/a}
\subsection{Setting the FRP Requirements}\label{sec3b}
\input{Section_III/b}
\subsection{Second Market Pass: Day-Ahead Market Clearing (DAMC) Problem}\label{sec3c}
\input{Section_III/c}
\section{Numerical Experiments}\label{sec4}
\input{Section_IV/4intro}
\subsection{Datasets and Benchmarks}\label{sec4a}
\input{Section_IV/4a}
\subsection{Results}\label{sec4b}
We next discuss the results.
\subsubsection{Case Study I}\label{sec4b1}
\input{Section_IV/4b_cs1}
\subsubsection{Case Study II Results}\label{sec4b2}
\input{Section_IV/4b_cs2}
\subsubsection{Case Study III Results}\label{sec4b3}
\input{Section_IV/4b_cs3}
\subsubsection{Case Study IV Results}\label{sec4b4}
\input{Section_IV/4b_cs4}
\section{Conclusion}\label{sec5}
\input{Section_V/5}
\appendices
\section{Nomenclature}\label{app_1}
\input{Appendices/nomenclature}
\vspace{-0.5cm}
\section{Stochastic Unit Commitment Problem Formulation}\label{app_2}
\input{Appendices/suc}
\vspace{-0.25cm}
\bibliographystyle{IEEEtran}
\bibliography{IEEEabrv,rampbib}
\balance
\vspace{-0.75cm}
\input{bios}
\end{document}

%% file: Section_0/packages.tex
\ifCLASSINFOpdf
\else
\fi
\usepackage{cite, authblk, amssymb, amsfonts, anyfontsize, soul, enumitem, amsmath, amsthm,algorithmic, graphicx, textcomp, xcolor, multicol, multirow, array, hhline, balance}
\usepackage[caption=false,font=footnotesize]{subfig}
\DeclareGraphicsExtensions{.png,.pdf}
\usepackage{algorithm, longtable, xtab, mathtools, stackengine, mathrsfs, setspace, bm, multirow, tikz, pgfplots, adjustbox, calc, url, xparse}
\usepackage{tikz}
\usetikzlibrary{positioning, shapes.geometric}
\definecolor{change_color}{HTML}{000000}
 \usepackage[draft]{changes}
\makeatletter
\@namedef{Changes@AuthorColor}{change_color}
\colorlet{Changes@Color}{change_color}
\makeatother
\definecolor{COLOR_INPUT}{HTML}{f9cf95}
\definecolor{COLOR_OUTPUT}{HTML}{47acb7}
\definecolor{COLOR_PROCESS}{HTML}{FA8072}
\definecolor{COLOR_EXCHANGE}{HTML}{a3d5db}

\definecolor{salmon}{HTML}{DC8072}
\definecolor{dblue}{HTML}{32738C}
\definecolor{lblue}{HTML}{4FAEC3}
\definecolor{lyellow}{HTML}{F7CF96}
\usepackage[colorlinks=true, citecolor=dblue, urlcolor= lblue, linkcolor=salmon]{hyperref}

%% file: Section_0/title.tex
\bstctlcite{IEEEexample:BSTcontrol}
\allowdisplaybreaks
\title{Flexible Ramping Product Procurement\\ in Day-Ahead Markets}
\author{Ogun~Yurdakul,~\IEEEmembership{Member,~IEEE,}%
	~Erik~Ela,~\IEEEmembership{Senior Member,~IEEE,}
	and~Farhad~Billimoria
\thanks{This work was supported in part by the U.S. Department of Energy Advanced Grid Modeling Program under Grant DE-OE0000875.}
\thanks{O. Yurdakul is with the Energy Systems and Infrastructure Analysis Division, Argonne National Laboratory, Lemont, IL 60439 USA (e-mail: oyurdakul@anl.gov).} 
\thanks{E. Ela is with Power Delivery and Utilization Department, Electric Power Research Institute, Palo Alto, CA 94304-1395 USA (e-mail: eela@epri.com).}
\thanks{F. Billimoria is with Energy \& Power Group, Department of Engineering Science, University of Oxford, Oxford OX1 3PJ, UK (e-mail: farhad.billimoria@wolfson.ox.ac.uk).}}
\markboth{Accepted for Publication in IEEE Transactions on Energy Markets, Policy and Regulation}%
{Accepted for Publication in IEEE Transactions on Energy Markets, Policy and Regulation}

%% file: Section_0/abstract_and_keywords.tex
\begin{abstract}
Flexible ramping products (FRPs) emerge as a promising instrument for addressing steep and uncertain ramping needs through market mechanisms. Initial implementations of FRPs in North American electricity markets, however, revealed several shortcomings in existing FRP designs. In many instances, FRP prices failed to signal the true value of ramping capacity, most notably evident in zero FRP prices observed in a myriad of periods during which the system was in acute need for rampable capacity. These periods were marked by scheduled but undeliverable FRPs, often calling for operator out-of-market actions. On top of that, the methods used for procuring FRPs have been primarily rule-based, lacking explicit economic underpinnings. In this paper, we put forth an alternative framework for FRP procurement, which seeks to set FRP requirements and schedule FRP awards such that the expected system operation cost is minimized. Using real-world data from U.S. ISOs, we showcase the relative merits of the framework in (i) reducing the total system operation cost, (ii) improving price formation, (iii) enhancing the the deliverability of FRP awards, and (iv) reducing the need for out-of-market actions.
\end{abstract}
 \begin{IEEEkeywords}
 day-ahead market, flexible ramping product, stochastic programming, unit commitment
 \end{IEEEkeywords}

%% file: Section_I/intro.tex
\IEEEPARstart{I}n many regions around the world, the share of renewables is approaching levels that would have been considered a far-fetched scenario roughly three decades ago, during the inception of organized wholesale electricity markets. Initially designed based on the economic and physical characteristics of dispatchable generators (DGs), markets and operations have been put under significant stress by the deepening penetration of renewables. Due to their highly volatile and uncertain power outputs, renewables oftentimes brought about tight supply conditions, increasing the frequency of price spikes in the real-time market (RTM). At the same time, system operators (SOs) noted their need to rely on ``[p]ersistent and systematic out-of-market actions” \cite{dame_final} so as to maintain system reliability.\par
SOs ascribe a majority of such tight supply conditions to a lack of ramping capability and scheduling inefficiencies rather than an underlying capacity shortage \cite{miso}. To address the need for flexibility through market mechanisms, SOs proposed several market design changes, which include most notably the introduction the flexible ramping product (FRP). Spearheaded by California ISO (CAISO) and Midcontinent ISO (MISO), the introduction of the FRP aimed to schedule sufficient ramping capability in the system so that the net load\footnote{While net load can take several definitions, in this article, we define net load as load minus utility-scale solar and wind generation.} that materializes in the RTM can be reliably met.\par
Now approximately six years after the initial introduction of the FRP, SOs have gained extensive insights into their effectiveness. Quite strikingly, SOs observed that the FRP hardly helped in reducing the need for manual interventions. CAISO observed that their manual interventions ``have remained high, and indeed increased, since the introduction of the FRP" \cite{caiso_dmm}. Indeed, FRPs were often stranded and not deliverable, as they were awarded to DGs located behind transmission constraints, even in cases where they were known in advance to be so. Such awards led to markedly low FRP prices even during periods of scarce ramping capability, failing to signal the true value of providing flexibility \cite{caiso_frp_ferc}. \par
Drawing on practical experience with current FRP designs, in this article, we set out to develop an alternative approach to FRP procurement. Our approach focuses on the day-ahead market (DAM), aligning with the SOs' recent objectives to extend the time horizon of the FRP and account for uncertainty over longer horizons \cite{caiso_dmm}. Most notably, we approach FRP procurement from a stochastic optimization lens, aiming to procure FRPs in a way that seeks to minimize the expected system operation cost under uncertain and variable net load.\par
We continue this article in Section \ref{sec2} by discussing the need for FRPs and reviewing the FRP procurement models deployed in practice and proposed in the literature. Accordingly, we lay out the specific contributions of our work in Section \ref{sec2}. In Section \ref{sec3}, we describe our FRP procurement methodology  and spell out its analytical underpinnings. Using real-world data harvested from U.S. ISOs, we conduct several case studies in Section \ref{sec4}, discuss the market outcomes obtained under the proposed methodology, and compare them against various benchmark methods. Finally, we present our concluding remarks and discuss policy implications in Section \ref{sec5}. 

%% file: Section_II/intro.tex
We begin this section by discussing the key factors that underpin the need for FRPs in power systems.

%% file: Section_II/a.tex
In the day-ahead time frame, the need for procuring ramping capacity is brought about by two principal factors: the variability and the uncertainty of net load. The former refers to the planned, anticipated change in net load across time periods. It encompasses both the change in net load between hourly intervals (inter-hourly variability) and that within the hour (intra-hourly variability). Since the net load can undergo precipitous climbs and drops within the hour, the ramping needs created by the intra-hourly variability of net load can be much more acute and steep with respect to those required to meet the inter-hourly variability of net load. This introduces additional complexity to the problem as the day-ahead decisions (in both the DAM and the reliability unit commitment process) are taken with an hourly granularity and thus fail to take account of the intra-hourly variability of net load. Ramping needs due to the uncertainty in net load refer to the fact that the exact net load levels that will eventuate cannot be known with perfect accuracy in advance and that SOs need to ensure that the unexpected ramping needs that may arise in real time can be reliably met. \par
To ensure that there is sufficient rampable capacity in the system, SOs have to frequently resort to out-of-market (OOM) actions, bringing DGs on-line by issuing exceptional commitment instructions. Too often, they leverage the tools at their disposal beyond what they are designed for. For instance, CAISO reported that it had to manually increase, on a routine basis, the load forecast used in the hour-ahead scheduling process so that additional ramping capacity could become available in real time \cite{ caiso_dmm}. \par
Clearly, a continuous need for such interventions points to a structural flaw in market design. As such, SOs set out to develop a novel market product that can (i) schedule sufficient ramping capability through market mechanisms, (ii) obviate the need for continuous market interventions and manual operator adjustments, and (iii) signal the true value of (and accordingly incent investments in) flexible ramping capability. 

%% file: Section_II/b.tex
The efforts to meet ramping needs through market mechanisms culminated in the introduction of the FRP in CAISO and MISO in 2010s, which was recently followed by the Southwest Power Pool (SPP). The chief objective of the FRP is to pre-position the DGs so as to schedule sufficient ramping capacity in the system. The procured upward and downward ramping capacity through up-FRP and down-FRP awards, respectively, can be deployed in real time if and when the net load changes beyond what can be met by the energy awards. \par
A salient aspect of the FRP is its pricing: in most applications, separate financial bids are not submitted for FRPs. Instead, FRP price is set based on opportunity costs that reflect the trade-off between the need for energy and ramping capacity. Specifically, the up-FRP (resp. down-FRP) price is set as the opportunity cost of the marginal resource that receives an up-FRP (resp. down-FRP) award and is equal to the difference between the variable cost of the marginal resource and the price of energy. Opportunity cost-based pricing ensures that the DGs that receive FRP awards can recoup the profits they forgo due to being held back from generating and not receiving energy awards. 
\par
SOs determine the up-FRP and down-FRP quantities that need to be procured by taking account of both the variability and uncertainty of net load.  To compute the ramping needs due to the uncertainty in net load, SOs draw on the percentiles of the historical net load forecast error distribution (such as the 2.5$^{\text{th}}$ and 97.5$^{\text{th}}$ percentiles in CAISO) \cite{ caiso_frp_ferc}. \par
\added{We next discuss the approaches adopted by different ISOs for FRP implementations. Constituting one of the earliest implementations, MISO introduced its FRP product (termed the ramp capability product) in 2016, which ensures sufficient 10-minute ramping capacity to cover for the forecasted net load variation in addition to net load uncertainty \cite{ychen_miso}. MISO procures FRPs both in the DAM and the RTM. In its market-clearing models, MISO enforces post-deployment transmission constraints, which aim to ensure that reserves can be delivered following a contingency event. These constraints rely on zonal sensitivities that approximate the impact of reserve deployment on transmission lines. While MISO does not enforce explicit zonal requirements for FRPs, the post-deployment transmission constraints could lead to different FRP prices across different zones \cite{miso_bpm}.}\par  
\added{In contrast to MISO, CAISO procures FRPs (termed flexible ramping product in CAISO) only in its RTM. Introduced in 2016, FRPs were procured to address both the variability and the uncertainty in net load \cite{caiso_frp_ferc}. However, the original FRP implementation in CAISO did not account for the nodal deliverability of the FRPs. More recently, SPP introduced its FRP product (termed ramp product) into its DAM and RTM on March 1, 2022 in order to address ten-minute ramping needs \cite{spp_q1_2023}. In all of these ISOs, FRPs are regarded as a distinct product and the provision of other ancillary services are not counted towards meeting FRP requirements. In fact, the existing market-clearing processes ensure that the FRPs compete for the same generator capacity with  regulation, spinning, and supplemental reserves \cite{miso_bpm}}.\par
The FRP has also garnered attention in the academic community. References \cite{hobbs,ela,rosenwald,spin} study FRP procurement in the RTM. In \cite{hobbs}, the authors set up a real-time unit commitment (UC) model with FRP requirements and compare their approach against a stochastic UC model. Among the several valuable insights presented in \cite{hobbs}, we especially take note of the finding, that absent the explicit consideration of energy costs in awarding FRPs, DGs ``with low commitment costs but high running costs might be overcommitted, increasing energy costs under high ramp outcomes". In \cite{ela}, the authors consider the ramping needs caused by the variability, but not the uncertainty, of net load and put forth an incentive-compatible pricing method for a multi-interval dispatch model. In contrast to \cite{hobbs,ela}, the approach suggested in \cite{rosenwald} sets both zonal and system-wide FRP requirements. The methodology laid out in \cite{spin} seeks to use a portion of the spinning reserve committed in the DAM to procure FRPs in the RTM. \par
FRP procurement in the DAM is studied in \cite{mujde,environment}, among which \cite{mujde} sets the FRP requirements by explicitly accounting for the ramping needs for each of the four 15-minute intervals within the hour. To determine the ramping needs caused by the uncertainty in net load, both \cite{mujde,environment} harness the 95\% confidence interval of net load, as adopted in CAISO and SPP. Both studies \cite{mujde,environment}, however, fail to take account of the ramping needs brought about by the difference between the cleared net demand and actual net load forecast of the SO. We expand upon the further shortcomings of existing methods in the next section.

%% file: Section_II/c.tex
Since the initial introduction of the FRP, the assessments of SOs laid out several shortcomings. A major concern raised by SOs is the deliverability of the procured FRPs. SOs noted that FRPs were at times stranded, as they were awarded to DGs located behind transmission constraints. A ramification of such FRP awards was that the price of FRP was markedly low, even in periods during which the system was in dire need for ramping capacity. This happened because it is more economical under existing market-clearing models to procure FRPs from DGs stranded behind transmission constraints as the opportunity cost of procuring FRPs in lieu of energy from such DGs is misleadingly low \cite{caiso_frp_ferc}. 
In MISO, the price of up-FRP equaled 0 \$/MW in 93.0\% of all dispatch intervals from April 1, 2017 to April 30, 2018, during some which there was an acute ramping need in the system \cite{cavicchi}. \added{Existing designs thus fail to signal the true opportunity cost of providing ramping capacity instead of energy, creating the false impression that ramping capacity is more abundant and cheaper than it actually is. In fact, in August 2022, CAISO submitted a tariff amendment to FERC to refine its FRP to address the deliverability issues observed in practice~\cite{caiso_frp_ferc}. As part of this tariff amendment, CAISO incorporated deployment scenarios to its market-clearing process, which serve to ensure that the deployment of the procured FRPs in either the upward or downward direction do not violate transmission constraints. Importantly, CAISO recognizes that this ``approach will not eliminate the occurrence of undeliverable flexible ramping product, but it will mitigate this risk… The CAISO’s approach will, however, avoid awarding flexible ramping product to resources that cannot, ex ante, reasonably be expected to deliver the awarded product if modeled uncertainty materializes''~\cite{caiso_frp_ferc}. Similarly, in SPP, the average up-FRP price was \$2.66/MW in the DAM and \$1.47/MW in the RTM during the fall of 2023, whereas the down-FRP prices have been consistently zero since its inception \cite{spp_q1_2023}.} At the same time, SOs noted that the need for OOM actions has not been reduced since the introduction of the FRP. Such inefficient OOM actions suppressed the RTM prices, further eating away at the revenue base of flexible DGs. \par
Another key shortcoming is that existing FRP procurement methods fail to sufficiently address the need for flexibility over longer time horizons, with CAISO procuring FRPs in only its RTMs. Further, current market designs cannot distinguish the energy costs of DGs when awarding reserve products. This is because, they lack an ex-ante assessment of the generation costs that will have to be borne if the procured reserve products are deployed. This is less of a concern for contingency reserves (which are rarely deployed) but more so for FRPs, which are called on frequently. \par
Despite their significance, the problems observed in practice have also been largely overlooked by the academic community. The deliverability of FRPs, for instance, are not addressed in any of the reviewed work from the literature. Except \cite{mujde,environment}, studies primarily focused on the RTM, failing to consider the procurement of FRPs to address the uncertainty and variability of net load in longer horizons. Reference \cite{hobbs} is the only study that describes the need for taking into account the generation costs that will result from the deployment of FRPs while determining the FRP awards. Nonetheless, \cite{hobbs} approximates these generation costs without considering the probability of deploying the procured FRPs.\par
Clearly, determining the quantity of FRPs that need to be procured plays a pivotal role in efficiently meeting net load ramps. Existing methods employed by SOs and proposed in the literature fail to address this problem from a stochastic optimization lens, resorting instead to rule-based methods (such as the 95\% confidence interval). Nevertheless, such one-size-fits-all methods lack an economic justification. In certain intervals, meeting the entire confidence interval (in lieu of the 95\%) of net load could bring forth a lower expected total cost. On the flip side, in some instances, meeting the 95\% confidence interval of net load could entail significant costs and it could be more rational to cover a tighter interval. \par
Motivated by these drawbacks, we here seek to develop an FRP procurement methodology for the DAM that can efficiently address the ramping needs due to both the variability and uncertainty in net load. Our key contributions include:
\begin{enumerate}
\item We propose a method for setting the FRP requirements such that the FRPs are procured only insofar as they help reduce the expected total operating cost of the system. As such, the imposed FRP requirement levels find explicit support in economic reasoning. The proposed market-clearing model comprises two market passes. The first market pass draws upon a stochastic UC model with a sub-hourly granularity, which explicitly assesses the uncertainty as well as the intra-hourly and inter-hourly variability of net load. The optimal decisions of the first market pass, however, are not financially binding and serve only as advisory decisions. Indeed, these advisory decisions are used in order to set the FRP requirements to levels that drive the DAM awards, as close as possible, to the optimal decisions of the first market pass. The second market pass harnesses a deterministic UC model formulated at an hourly granularity, whereby energy and FRP are co-optimized to determine financially-binding DAM energy and FRP awards. As such, our methodology adheres to the salient characteristics of today’s DAMs.
\item Our approach sees to it that the procured FRPs are deliverable. Here, our objective is \textit{not} to schedule the FRP awards such that the FRPs can be delivered for any possible ramping need. Instead, we ensure that the DGs, that are verified before the fact to be able to deliver ramping capability, are prioritized in receiving FRP awards. Through case studies conducted using real-world data, we show that such an approach leads to improved price formation, yielding non-zero FRP prices commensurate with the scarcity of ramping capacity.
\item In determining the FRP awards, the proposed methodology takes into account, ex-ante, the generation costs that may arise with the deployment of FRPs. This design choice imparts our methodology the capability to fend off unexpectedly high costs under steep ramping needs.
\end{enumerate}
At the time of writing this publication, SOs mapped out certain market design changes that mirror our contributions. For instance, as part of its DAM enhancements effort \cite{dame_final}, CAISO proposes the introduction of reliability capacity and imbalance reserve products, which seek to address the ramping needs driven by the variability and uncertainty of net load, respectively, in the day-ahead time frame. \added{Similarly, in the Winter 2024 State of the Market Report, the Market Monitoring Unit of SPP reiterated its outstanding recommendation on improving FRP outcomes with respect to stranded FRP awards, stating that addressing this issue could change pricing outcomes \cite{spp_2024_q1}.} Such cases not only drive home how acute of a need it is to address the identified gaps, but also reaffirm the practical relevance of our contributions. In the next section, we lay out our contributions in detail by taking a deep dive into our methodology. 

%% file: Section_III/intro.tex
In this section, we describe our proposed methodology, which co-optimizes energy and FRP awards in the DAM. The proposed methodology is composed of two market passes, each with a different temporal granularity. To clearly present the salient characteristics of each market pass, we begin by describing our time notation. The nomenclature and the detailed problem formulations are furnished in Appendix~\ref{app_1} and Appendix~\ref{app_2}, respectively.\par
In our problem setup, we take one day as our scheduling horizon. We denote by $h$ the index for each hour of a day and construct the set $\mathscr{H} \coloneqq \{ h \colon h=1,\ldots,24\}$. We decompose each hour into $K$ intra-hourly sub-periods, each with an equal duration of $\zeta$ minutes. For each hour $h\in \mathscr{H}$, we construct the set of sub-periods as $\mathscr{K}_h \coloneqq \{k \colon k=(h-1)\times K + 1,\ldots,h\times K\}$, where $k$ denotes the index for each intra-hourly sub-period. We define the set of all sub-periods over the scheduling horizon by $\mathscr{K}\coloneqq\cup_{h\in\mathscr{H}} \mathscr{K}_h$. \par 
The notation used in this article expressly distinguishes the temporal granularity of the variables and parameters. We denote by $x[k]$ the variable $x$ in sub-period $k$, which adopts $\zeta$ minutes as the smallest indecomposable unit of time. Conversely, the term $x(h)$ denotes the variable $x$ in hour $h$, which has an hourly granularity, ignoring any phenomena that occur within the hour. We next describe each market pass in turn.

%% file: Section_III/a.tex
In the first market pass, our primary objective is to provide advisory (i.e., financially non-binding) recommendations on the DAM energy and FRP awards. To form these recommendations, we seek answers to the following questions: (i) How should the FRP awards be scheduled such that their delivery is reasonably ensured, if or when they are called upon? (ii) Which DGs should receive FRP awards so that the deployment of the awarded FRPs will not lead to excessive production costs? (iii) What are prudential up-FRP and down-FRP requirements that will guarantee meeting ramping needs only to the extent that the expected total cost of the system is reduced? Here, the latter question calls for striking a balance between setting overly conservative requirements for FRPs, some of which will be rarely called upon (thus commanding hefty generator costs), and under-hedging the risk of not meeting ramping needs (which entails regular OOM actions and/or load curtailment, commanding hefty penalty costs). Above all, such FRP requirements must not be laid down on an ad-hoc basis and must instead be underpinned by an economic rationale.\par
As it relates to the UC problem, two-stage stochastic optimization models seek to determine the here-and-now commitment decisions and the wait-and-see dispatch decisions that minimize the expected total cost of the system. This suggests a natural applicability to the first market pass, as a stochastic unit commitment (SUC) model that includes transmission constraints and is formulated at a sub-hourly granularity inherently provides justifiable answers to all of the above questions. As such, we employ an SUC model in the first market pass.\par
In our formulation, we model the uncertain net load using a set of scenarios. The term $\xi^{\omega}$ denotes the net load across each transmission bus over the intra-hourly time periods of a day in scenario $\omega \in \Omega$, and the term $\pi^{\omega}$ denotes the probability of the scenario $\omega \in \Omega$. The first-stage problem of the first market pass is succinctly stated as

{
\begin{IEEEeqnarray}{ll}
\underset{\underset{w_g[k]}{u_g[k], v_g[k]} }{\text{\fontsize{9}{11}\selectfont minimize}} &\hspace{0.3cm}  \sum_{k \in \mathscr{K}} \sum_{g \in \mathscr{G}} \Big[ \alpha^{u}_{g} u_{g}[k] + \alpha^{v}_{g} v_{g}[k] \Big]   \nonumber\\
& \hspace{1.5cm} + \sum_{\omega \in \Omega} \pi^{\omega} \mathcal{Q}({x},{\xi^{\omega}}), \label{obj}\\
 \text{\fontsize{9}{11}\selectfont subject to}&\hspace{0.3cm} u_g[k], v_g[k], w_g[k] \in \mathscr{X},\\ 
&\hspace{0.3cm}u_{g}[k] = u_{g}[K\times h],\nonumber \\
&\hspace{1.2cm}\forall k \in \mathscr{K}_h\setminus \{K\times h\}, \forall h \in \mathscr{H}, \label{int}
\end{IEEEeqnarray}}
where $x$ is the vector of first-stage decisions comprising the binary commitment $u_g[k]$, startup $v_g[k]$, and shutdown $w_g[k]$ variables of all DGs $g \in \mathscr{G}$ over all time periods $k \in \mathscr{K}$. The objective \eqref{obj} of the first-stage problem is to minimize the sum of the commitment $\alpha^{u}_{g} u_{g}[k]$ and start-up costs $\alpha^{v}_{g} v_{g}[k]$ over all DGs and time periods plus the expected dispatch and load curtailment costs $\sum_{\omega \in \Omega} \pi^{\omega} \mathcal{Q}({x},{\xi^{\omega}})$. The set $\mathscr{X}$ denotes the feasibility region of the first-stage decisions, characterized by the minimum uptime and downtime constraints of the DGs and the logical constraints that relate the commitment, startup, and shutdown variables. 
We enforce in \eqref{int} that the commitment decisions do not change within the hour $h \in \mathscr{H}$. The constraint \eqref{int} makes sure that the first market pass decisions can be implemented in the second market pass, which, as shall be spelled out below, has an hourly granularity and thus cannot support commitment instructions that vary within the hour.\par
For a specific choice of first-stage decisions $x$ and a net load realization $\xi^{\omega}$, $\mathcal{Q}(x, \xi^{\omega})$ is the optimal value of the following second-stage problem
\vspace{-0.5cm}

{
\begin{IEEEeqnarray}{ll}
\underset{\begin{subarray}{c}p_g[k], p_g^{s}[k], \\p_{n}^c[k]\end{subarray}}{\text{minimize}}& \hspace{0.6cm} \sum_{k \in \mathscr{K}}\bigg[ \sum_{g \in \mathscr{G}} \sum_{s \in \mathscr{S}_g} {\alpha}^{s}_{g}p^{s}_{g}[k] + \sum_{n \in \mathscr{N}} \alpha^{c}p^{c}_{n}[k]\bigg], \label{obj2}\\
\text{subject to}& \hspace{0.6cm} p_g[k], p_g^{s}[k], p_{n}^c[k],\,p^{net}_{n}[k] \in \mathscr{Y}(x, \xi^{\omega}),\label{sec_stage}\\
& \hspace{0.6cm} p^{net}_{n}[k] = \sum_{g \in \mathscr{G}_{n}} p_{g}[k] + p^{c}_{n}[k] - {\xi}^{\omega}_{n}[k],\nonumber \\
& \hfill \forall n \in \mathscr{N}, \label{netpower}\\
& \hspace{0.6cm} \sum_{n \in \mathscr{N}} p^{net}_{n}[k]  = 0, \label{pb}\\
& \hspace{0.6cm} \underline{f}_{\ell} \leq \sum_{n \in \mathscr{N}}  \Psi^{\ell}_{n}p^{net}_{n}[k] \leq \overline{f}_{\ell}\nonumber \\
& \hfill \forall \ell \in \mathscr{L}, \label{pf}
\end{IEEEeqnarray}}
where \eqref{netpower}--\eqref{pf} hold for all $k \in \mathscr{K}$. We represent the decision variables of the second-stage problem \eqref{obj2}--\eqref{pf} by $y$, which are comprised of the dispatch level $p_g[k]$ of each DG $g \in \mathscr{G}$, the power generated from each segment $p_g^{s}[k]$ of the piecewise linear production cost segments of DG $g \in \mathscr{G}$, the net power injection $p^{net}_{n}[k]$ at transmission system bus $n \in \mathscr{N}$, and the curtailed load $p^{c}_{n}[k]$ at bus $n \in \mathscr{N}$ for each period $k \in \mathscr{K}$. The objective \eqref{obj2} of the second-stage problem is to minimize the total power dispatch cost plus the penalty cost due to load curtailment. The constraints on the power generation and ramping limits are encapsulated in \eqref{sec_stage}, which further includes constraints on the power dispatch from each segment of the piecewise linear cost function of DGs. We express the net real power injection at each node $n \in \mathscr{N}$ in \eqref{netpower} with the convention that $p^{net}_{n}[k]>0$ if real power is injected into the system. The term $\xi^\omega_{n}[k]$ in \eqref{netpower} denotes the net load realization in scenario $\omega \in \Omega$ at bus $n \in \mathscr{N}$ in period $k \in \mathscr{K}$. We state the system-wide power balance constraint in \eqref{pb}. We use the DC power flow model to state the transmission constraints and utilize injection shift factors (ISFs) for network representation \cite{vanhorn}. Finally, in \eqref{pf}, we express the real power flow on each transmission line $\ell \in \mathscr{L}$ in terms of nodal injections and {ISF}s and constrain it to be within its line flow limits.\par
The optimal SUC solution affords critical information on the rampable capacity required for minimizing the expected total system operation cost. Specifically, the optimal SUC solution reveals the \textit{sweet spot} for the trade-off between the generator costs incurred by meeting the ramping needs and the penalty costs incurred by not meeting said needs and instead resorting to load curtailment. By weighing the net load realization and the ramping need in each scenario against its probability, the SUC solution further provides the optimal DG schedules for satisfying the necessary ramping needs with least expected total cost. In determining the DG schedules, it explicitly assesses the requisite startup costs in conjunction with the expected production costs---all while verifying the deliverability of the awarded schedules by observing the transmission constraints.\par
As identified previously in \cite{hobbs}, if it were possible to commit DGs based on the optimal SUC decisions, it would not be necessary to separately acquire FRPs. 
In practice, however, it is not possible to employ the SUC model to clear the DAM on at least two accounts. First is the computational burden of solving a large-scale stochastic optimization problem where there are stringent time limits on publishing the market outcomes. The second is that it would necessitate changes to virtually every regulation and instrument in today’s markets. These notwithstanding, the optimal SUC solution can be effectively exploited as best practice guidelines in setting the FRP requirements. To that end, SOs can tailor the SUC formulation to its requirements and computational capabilities by, for instance, adjusting the number of scenarios or the modeled transmission constraints. At the same time, SOs can solve the SUC problem ahead of time using a set of scenarios that are representative of, not only a single day, but rather of a group of comparable days, thus obviating the need to solve SUC on a daily basis. In the next section, we spell out how we use the optimal SUC solution to set the FRP requirements.

%% file: Section_III/b.tex
In the previous subsection, we laid out how the optimal SUC solution meets the ramping needs faced under each scenario only as long as doing so lends itself to reducing the expected total system operation cost. For instance, assume that the steepest upward ramping need between time periods $k$ and $k+1$ is observed in scenario $\omega \in \Omega$, requiring a rampable capacity of $\sum_{n \in \mathscr{N}} \big(\xi^{\omega}_{n}[k+1] - \xi^{\omega}_{n}[k]\big)$ MW. Assume further that the optimal second-stage solution to the SUC problem for scenario $\omega $, denoted by $y^{*}_{\omega}$, elects to schedule only a part of the requisite ramping capacity and curtails some load in periods $k$ and $k+1$, thus satisfying the following ramping need

\begin{IEEEeqnarray}{l}
\sum_{n \in \mathscr{N}} \bigg(\Big[\xi^{\omega}_{n}[k+1] -p_{n}^{c}[k+1]\big|_{y_{\omega} = y^{*}_{\omega}}\Big] \nonumber \\
\hfill - \Big[\xi^{\omega}_{n}[k] -p_{n}^{c}[k]\big|_{y_{\omega} = y^{*}_{\omega}}\Big]\bigg).\label{ramp_example}
\end{IEEEeqnarray}
In this case, it is warranted and needed to procure FRPs such that the ramping need in \eqref{ramp_example} is satisfied. On the flip side, procuring FRPs beyond \eqref{ramp_example} entails excess generation costs that surpass the reduction in the expected cost of load curtailment. \par
In setting the FRP requirements, it is critical to recognize that the DAM clears at an hourly granularity, which is coarser than the sub-hourly granularity of the ramping needs identified by the optimal SUC solution. The FRP requirements imposed in the DAM need to ensure that the procured FRPs can meet the ramping needs that are, under the optimal SUC solution, elected to be met between \textit{all} intra-hourly sub-periods within each hour. In this light, we capitalize on the optimal SUC solution to set the up-FRP requirements, denoted by $\underline{\rho}^{\uparrow}(h)$, and the down-FRP requirements, denoted by $\underline{\rho}^{\downarrow}(h)$, as

{\
\begin{IEEEeqnarray}{ll}
\underline{\rho}^{\uparrow}(h)  =& K \times\bigg( \quad\underset{k \in \mathscr{K}_h}\max\quad\underset{\omega \in \Omega}\max\,\sum_{n\in\mathscr{N}} \bigg[\Big[\xi^{\omega}_{n}[k+1] \nonumber \\
&\hspace{-0.5cm}-p_{n}^c[k+1]\big|_{{y}_{\omega} = y^{*}_{\omega}}\Big] - \Big[\xi^{\omega}_{n}[k] - p_{n}^c[k]\big|_{{y}_{\omega} = y^{*}_{\omega}}\Big]\bigg),\label{upfrp}\\
\underline{\rho}^{\downarrow}(h)  =&-K \times\bigg( \quad\underset{k \in \mathscr{K}_h}\min\quad\underset{\omega \in \Omega}\min\,\sum_{n\in\mathscr{N}} \bigg[\Big[\xi^{\omega}_{n}[k+1] \nonumber \\
&\hspace{-0.5cm}-p_{n}^c[k+1]\big|_{{y}_{\omega} = y^{*}_{\omega}}\Big] - \Big[\xi^{\omega}_{n}[k] - p_{n}^c[k]\big|_{{y}_{\omega} = y^{*}_{i}}\Big]\bigg).\label{dwfrp}
\end{IEEEeqnarray}}
Put into words, \eqref{upfrp} and \eqref{dwfrp} begin by identifying the highest satisfied upward and downward ramping need, respectively, over all scenarios for each sub-period $k \in \mathscr{K}$. Next, for each hour $h \in \mathscr{H}$, they compare the identified needs of the child sub-hourly periods and determine the highest ramping need satisfied within each hour. Finally, the obtained intra-hourly ramping needs are multiplied by $K$ so as to ensure that the hourly requirements can provide the desired ramping capability across all intra-hourly time periods in $\zeta$-minutes.\par
The astute reader will recognize that while we had set out to answer three principal questions in the beginning of this section, setting FRP requirements through \eqref{upfrp} and \eqref{dwfrp} neither ensures the deliverability of the procured FRPs nor does it avoid high production costs that may be incurred with the deployment of the procured FRPs. In the next section, we lay out how we draw upon the optimal first-stage solution to the SUC problem, i.e., $x^{*}$, in addressing these issues.

%% file: Section_III/c.tex
We next delineate the day-ahead market clearing (DAMC) problem, which is executed as the second pass of the proposed methodology to co-optimize energy and FRP schedules and to issue financially binding DAM awards. To ensure compliance with existing DAM designs adopted in the U.S., we construct the DAMC problem as a deterministic UC problem formulated at an hourly granularity. Most notably, we capitalize on the up-FRP and down-FRP quantities derived in the previous section to size the up-FRP and down-FRP requirements in the DAM. The DAMC problem is formulated as

{
\begin{IEEEeqnarray}{ll}
\underset{\begin{subarray}{c}u_g(h), v_g(h), p_g(h),\\ p_g^{s}(h), p_{n}^{c}(h), d_{n}(h), \\ r^{\uparrow}_{sf}(h), r^{\downarrow}_{sf}(h)\end{subarray}}{\text{minimize}} &\hspace{0.2cm} \sum_{h \in \mathscr{H}}\bigg[ \sum_{g \in \mathscr{G}} \Big[ \alpha^{u}_{g} u_{g}(h)+ \alpha^{v}_{g} v_{g}(h)  \nonumber \\
&\hspace{0.2cm} + \sum_{s \in \mathscr{S}_g} {\alpha}^{s}_{g}p^{s}_{g}(h)\Big]\nonumber\\[8pt]
 &\hspace{0.2cm} + \sum_{n \in \mathscr{N}} \alpha^{c}p_{n}^{c}(h)  \nonumber\\[8pt]
  &\hspace{0.2cm} + \,\alpha^{r} \Big(r^{\uparrow}_{sf}(h) + r^{\downarrow}_{sf}(h)\Big)\bigg], \label{damcobj}
\end{IEEEeqnarray}}
{\begin{IEEEeqnarray}{l}   
\text{\selectfont subject to}\nonumber\\[10pt]
\text{(i) logical constraints relating the binary variables:} \nonumber\\
u_{g}(h)-u_{g}(h-1) = v_{g}(h) - w_{g}(h), \label{damcbin}\\
u_{g}(h)-u_{g}^{\circ} = v_{g}(h) - w_{g}(h), \label{damcst}\\
u_{g}(h), v_{g}(h), w_{g}(h)  \in \{0,1\},\label{damcbinreq}\\[10pt]
\text{(ii) minimum uptime and downtime constraints:} \nonumber\\
\sum_{h'=h-T^{\uparrow}_{g}+1}^{h} v_{g}(h') \leq u_{g}(h) \hspace{1.1cm} \forall h \in \{T^{\uparrow}_{g},\ldots,24\},\label{damcupt}  \\
\sum_{h'=1}^{\min\{u_g^{\circ}\times(T^{\uparrow}_{g} - T^{\uparrow, \circ}_{g}), \,24\}} w_{g}(h') = 0, \label{damciupt}  \\
\sum_{h'=h-T^{\downarrow}_{g}+1}^{h} w_{g}(h') \leq1- u_{g}(h) \hspace{.4cm}  \forall h \in \{T^{\downarrow}_{g},\ldots,24\}, \label{damcdwt}  \\
\sum_{h'=1}^{\min\{(1-u_g^{\circ})\times(T^{\downarrow}_{g} - T^{\downarrow, \circ}_{g}), \,24\}} v_{g}(h') = 0,\label{damcidwt}  \\[10pt]
\text{\selectfont (iii) constraints on power generation levels:} \nonumber\\
0 \leq p_{g}(h) \leq (\overline{P}_{g}-\underline{P}_{g})u_{g}(h) ,\label{damcgl} \\
p_g(h) \leq p_{g}^{\circ}+\Delta_g^{\uparrow} u_{g}^{\circ}+(\Delta_g^{\uparrow, 0} -\underline{P}_g) v_g(h),\label{damcrul1} \\
p_g(h) \geq p_{g}^{\circ}-\Delta_g^{\downarrow} u_{g}^{\circ}+\Big(\Delta_g^{\downarrow}-p_{g}^{\circ}\Big) w_g(h),\label{damcrll1}\\
p_g(h) \leq p_g(h-1)+\Delta_g^{\uparrow} u_g(h-1) \nonumber \\
\hspace{1.5cm}+(\Delta_g^{\uparrow, 0} -\underline{P}_g) v_g(h)\label{damcrul} \\
p_g(h) \geq p_g(h-1)-\Delta_g^{\downarrow} u_g(h-1)\nonumber \\
\hspace{1.5cm}+\Big(\Delta_g^{\downarrow}-p_g(h-1)\Big) w_g(h),\label{damcrll}\\
p_g(h) \leq w_g(h+1) (\Delta^{\downarrow, 0}_g - \underline{P}_g)\nonumber \\
\hspace{1.5cm}+\Big(1-w_g(h+1)\Big)\left(\overline{P}_g-\underline{P}_g\right),\label{damcsd}\\
p_{g}(h) = \sum_{s \in \mathscr{S}_g} p_{g}^{s}(h) ,\label{damcls}\\
0 \leq p^{s}_{g}(h) \leq \overline{P}^{s}_{g} - \overline{P}^{\,s-1}_{g} \hspace{2.7cm} \forall {s} \in \mathscr{S}_g,\label{damclsl} \\[10pt]
\text{\selectfont (iv) nodal power balance and line flow constraints:} \nonumber\\
p_{n}^{net}(h) = \sum_{g \in \mathscr{G}_{n}} p_{g}(h) + p_{n}^{c}(h) - d_{n}(h) \hspace{0.6cm} \forall n \in \mathscr{N},\label{damcnetpower} \\
\sum_{n \in \mathscr{N}} p_{n}^{net}(h)  = 0, \label{damcpb}\\
d_{n}(h) = \hat{d}_{n}(h)  \quad \xleftrightarrow{\hspace{.5cm}} \quad \lambda_{n}(h) \hspace{1.6cm}  \forall n \in \mathscr{N},  \label{damclmp}\\
\underline{f}_{\ell} \leq \sum_{n \in \mathscr{N}}  \Psi_{n}^{\ell}p_{n}^{net}(h) \leq \overline{f}_{\ell} \hspace{2.45cm} \forall \ell \in \mathscr{L}, \label{damcpf} \\
p_{n}^c(h) \geq 0 \hspace{4.6cm} \forall n \in \mathscr{N},\label{damcnn3}\\[10pt]
\hspace{0.0cm}\text{(v) FRP constraints for DGs:} \nonumber\\
-\Delta_g^{\downarrow} u_g(h)+(\Delta_g^{\downarrow}-\Delta_g^{\downarrow, 0}) w_g(h+1)+\underline{P}_g v_g(h+1) \nonumber \\
\hspace{1.5cm} \leq r_g^{\uparrow}(h)   \nonumber \\
\hspace{1.5cm} \leq \Delta_{g}^{\uparrow} u_g(h+1) + \Big(\Delta_{g}^{\uparrow, 0} - \Delta_{g}^{\uparrow}\Big) v_g(h+1),\label{damcupfrplimit1}\\
r_g^{\uparrow}(h) \leq \overline{P}_g u_g(h+1) - \underline{P}_g u_g(h) ,\label{damcupfrplimit2}\\
-\Delta_g^{\uparrow} u_g(h+1)+(\Delta_g^{\uparrow}-\Delta_g^{\uparrow, 0}) v_g(h+1)\nonumber \\
\hspace{1.5cm} \leq r_g^{\downarrow}(h) \nonumber \\ 
\hspace{1.5cm} \leq \Delta_{g}^{\downarrow} u_g(h) + \Big(\Delta_{g}^{\downarrow, 0} - \Delta_{g}^{\downarrow}\Big) w_g(h+1) \nonumber \\ 
\hspace{1.85cm} - \underline{P}_g v_g(h+1), \label{damcdwfrplimit1}\\
r_g^{\downarrow}(h) \geq - \overline{P}_g u_g(h+1) + \underline{P}_g u_g(h),\label{damcdwfrplimit2}\\
-\underline{P}_g+\underline{P}_g u_g(h+1) \nonumber \\
\hspace{1.5cm} \leq r^{\uparrow}_g(h)+p_g(h) \nonumber \\
\hspace{1.5cm} \leq \overline{P}_g-\underline{P}_g u_g(h)+(\Delta_g^{\uparrow, 0} - \overline{P}_g) v_g(h+1),\label{damcuprfrp_and_dispatch_1} \\
r_g^{\uparrow}(h)+p_g(h)  \nonumber \\
\hspace{1.5cm} \leq \Delta_{g}^{\downarrow, 0}w_g(h+2) + \overline{P}_g \big(1-w_g(h+2)\big), \label{damcuprfrp_and_dispatch_2}\\
 -\underline{P}_g+\underline{P}_g u_g(h+1) \nonumber \\
\hspace{1.5cm}\leq -r^{\downarrow}_g(h)+p_g(h)  \nonumber \\
\hspace{1.5cm}\leq \overline{P}_g-\underline{P}_g u_g(h)+(\Delta_g^{\uparrow, 0} - \overline{P}_g) v_g(h+1),\label{damcdwrfrp_and_dispatch_1}\\ 
-r_g^{\downarrow}(h)+p_g(h) \nonumber \\
\hspace{1.5cm}\leq \Delta_{g}^{\downarrow, 0}w_g(h+2) + \overline{P}_g \big(1-w_g(h+2)\big),\label{damcdwrfrp_and_dispatch_2}\\[10pt]
\text{(vi) system-wide up-FRP and down-FRP requirements:}\nonumber\\
\sum_{g \in \mathscr{G}} r^{\uparrow}_g(h) + r^{\uparrow}_{sf}(h) \geq  \underline{\rho}^{\uparrow}(h) \quad \xleftrightarrow{\hspace{.5cm}} \quad \varphi^{\uparrow}(h),\label{damcupfrp}\\
\sum_{g \in \mathscr{G}} r^{\downarrow}_g(h) + r^{\downarrow}_{sf}(h) \geq  \underline{\rho}^{\downarrow}(h) \quad \xleftrightarrow{\hspace{.5cm}} \quad \varphi^{\downarrow}(h),\label{damcdwfrp}
\end{IEEEeqnarray}}
where \eqref{damcbinreq}--\eqref{damcgl} and \eqref{damcls}--\eqref{damcdwfrp} hold for all hours $h \in \mathscr{H}$, \eqref{damcst}, \eqref{damcrul1}, and \eqref{damcrll1} hold for $h \in \{1\}$, and \eqref{damcbin}, \eqref{damcrul}, and \eqref{damcrll} hold for $h \in \mathscr{H} \setminus \{1\}$. The objective of the DAMC problem \eqref{damcobj} is to minimize the commitment costs $\alpha^{u}_{g} u_{g}(h)$, startup costs $\alpha^{v}_{g} v_{g}(h)$, and dispatch costs $\sum_{s \in \mathscr{S}_g} {\alpha}^{s}_{g}p^{s}_{g}(h)$ of all DGs plus the penalty costs due to involuntary load curtailment $\sum_{n \in \mathscr{N}} \alpha^{c}p_{n}^{c}(h)$ and FRP shortfall $\alpha^{r} \big(r^{\uparrow}_{sf}(h) + r^{\downarrow}_{sf}(h)\big)$. The terms $r^{\uparrow}_{sf}(h)$ and $r^{\downarrow}_{sf}(h)$ represent the shortfall in meeting the system-wide up-FRP and down-FRP requirements, respectively, and $\alpha^{r}$ is the penalty cost for FRP shortfall. The notation used in the publication is provided in Appendix~\ref{app_1}.\par
The constraints on the FRPs that a DG can be cleared for are shaped by the DG’s commitment, startup, shutdown, and dispatch decisions as well as its ramping limits. We set forth these constraints in \eqref{damcupfrplimit1}--\eqref{damcdwrfrp_and_dispatch_2} based on the formulation laid out in \cite{hobbs}. \added{Constraints \eqref{damcupfrplimit1}--\eqref{damcupfrplimit2} state the lower and upper bounds on the up-FRP that can be provided by a DG based on its ramping limits and the commitment statuses in the current and next time periods. Analogously, \eqref{damcdwfrplimit1}--\eqref{damcdwfrplimit2} limit the down-FRP provision based on the commitment schedules and ramping limits. We state in \eqref{damcuprfrp_and_dispatch_1}--\eqref{damcuprfrp_and_dispatch_2} the shared limits on the energy and up-FRP awards, which illustrate how the energy awards compete for the same capacity with the up-FRP awards. The analogues of these constraints for the down-FRP awards are stated in \eqref{damcdwrfrp_and_dispatch_1}--\eqref{damcdwrfrp_and_dispatch_2}.} In the DAMC formulation, we recognize that the \textit{bid-in} net demand can deviate from the net load forecasted by the SO. In this light, we distinguish in our notation the bid-in net demand at bus $n$, which is notated as $\hat{d}_{n}$, from the net load in scenario $\omega \in \Omega$ at bus $n \in \mathscr{N}$, which is notated as $\xi_{n}^{\omega}$. We assume that the net demand is perfectly inelastic and has an infinite willingness to pay.\par
\added{In line with existing practices of U.S. ISOs (see \cite{frp_app_b} for CAISO and \cite[p. 222]{miso_bpm} for MISO), we set the system-wide up-FRP price $\varphi^{\uparrow}(h)$ and the down-FRP price $\varphi^{\downarrow}(h)$ based on the the shadow price of the up-FRP \eqref{damcupfrp} and down-FRP \eqref{damcdwfrp} requirement constraints, respectively. We further draw upon the dual variable of \eqref{damclmp} to set the LMP $\lambda_{n}(h)$ at each bus $n \in \mathscr{N}$ in hour $h \in \mathscr{H}$. Put into words, we initially define the net demand at node $n$ as a variable through $d_n$, and we subsequently set it equal to the parameter $\hat{d}_{n}$ in \eqref{damclmp}. The dual variable $\lambda_{n}(h)$ of \eqref{damclmp} signifies the change in the objective value \eqref{damcobj} with a slight increase or decrease in the net demand at node $n$, which directly translates into the definition of the LMP \cite{m39rep}.}\par
\added{A remark is in order on the modeling choices adopted in the proposed framework. Observe that the DAMC problem omits key elements included in the actual DAM models solved by ISOs, such as operating reserves, virtual bids, and contingencies. In addition, the ISOs draw upon several processes between DAM clearing and delivery, such as reliability unit commitment, look-ahead unit commitment, and hour-ahead scheduling activities, executed both before and during the operating day. We chose not to model such activities as doing so would overcomplicate the proposed approach without shedding light on the fundamental issues surrounding FRPs.}\par
The DAMC problem enforces the system-wide up-FRP and down-FRP requirements through \eqref{damcupfrp} and \eqref{damcdwfrp}, respectively. A major concern that arises from procuring FRPs based solely on \eqref{damcupfrp} and \eqref{damcdwfrp} is that such a procurement scheme falls short of considering the production costs and the locations of DGs while determining the FRP awards. This runs counter to our objective of ensuring the deliverability of FRPs and mitigating excessive production costs that may arise with FRP deployment. We explain this potential counter-intuitive outcome using a toy example.\par
\begin{figure}[ht]
\includegraphics[width=\linewidth]{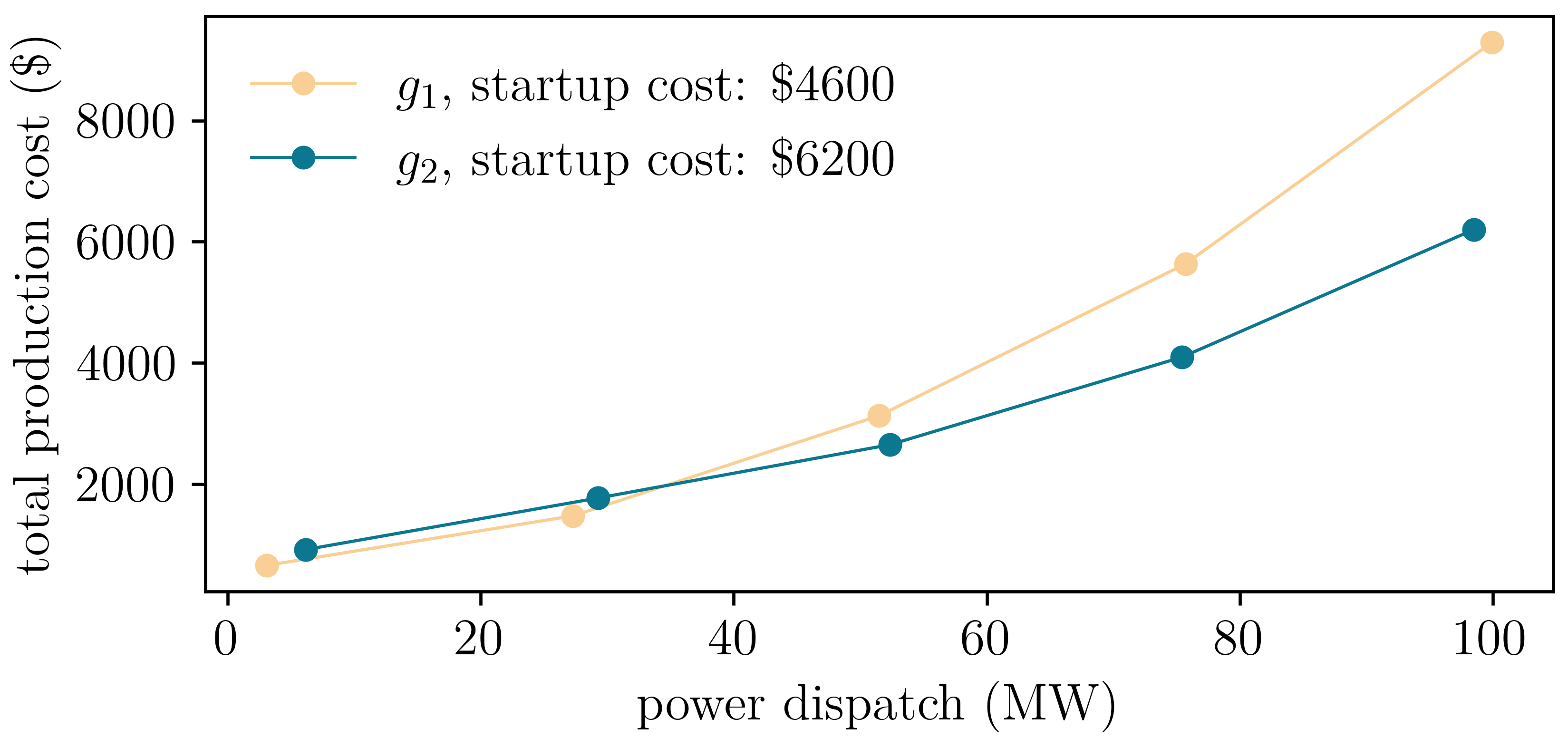}
\caption{Cost curves of DGs in the toy example}
\label{toy}
\end{figure}

Suppose the up-FRP requirement in hour $h$ is set at $\underline{\rho}^{\uparrow}(h) = 80$ MW/h and there are two DGs, $g_1$ and $g_2$, which are currently offline and are competing for the up-FRP awards. Suppose further that each DG can meet the entire up-FRP requirement on its own. The cost curves of the two DGs are depicted in Fig \ref{toy}. \par
\input{Section_III/flowchart}
In this example, scheduling the up-FRP award solely through \eqref{damcupfrp} will favor $g_1$ for the up-FRP award, as the objective function \eqref{damcobj} considers only the fixed commitment and startup costs of the DGs in scheduling FRP awards, thus always giving a higher precedence to the DG with the lower commitment and startup cost. In practice, however, the optimal FRP awards that minimize the expected total system operation cost hinge on the quantity and the probability of the ramping needs that may materialize. If a large portion of the awarded FRPs is expected to be deployed frequently, it may make more sense to award up-FRPs to $g_2$, as the resulting lower production costs (compared to that under $g_1$) will outweigh its higher startup costs. Conversely, if the awarded FRP is expected to be deployed only to meet tail events, awarding up-FRP to $g_1$ may yield a lower expected total cost. Furthermore, procuring FRPs through \eqref{damcupfrp} and \eqref{damcdwfrp} completely overlooks the deliverability issues. In the event that $g_1$ has a positive shift factor to a binding transmission constraint, the FRP awarded to $g_1$ may fail to get deployed.  \par
The SUC solution, however, innately addresses both of these problems. Carrying on with the above example, the optimal SUC solution pinpoints whether $g_1$ or $g_2$ needs to receive the up-FRP award so that the expected total cost of the system is minimized. This is ensured by SUC's ability to comprehensively assess the startup and commitment costs in conjunction with the production costs of the DGs, while evaluating the uncertain ramping needs against their probabilities. On top of that, the SUC problem determines optimal schedules by taking explicit account of the transmission constraints, thus verifying in advance that the scheduled ramping capacity can be delivered in each scenario. \added{As a result, by addressing the net load realizations and ramping needs in all scenarios without violating the transmission constraints, the SUC problem serves a similar purpose as the post-deployment transmission constraints of MISO and the deployment scenarios of CAISO.} On these grounds, we seek to leverage the optimal commitment decisions of the SUC problem in procuring FRPs in the DAMC formulation. Specifically, our objective is to ensure that the DGs that are committed under the optimal SUC solution remain committed in the optimal DAMC solution, thereby clearing for DAM energy and FRP awards with a higher priority. For this purpose, we incorporate the following constraint into the DAMC problem
\begin{IEEEeqnarray}{l}
u_g(h) \geq u_{g}[K\times h]\big|_{x=x^{*}}\quad \forall g \in \mathscr{G}, \forall h \in \mathscr{H}. \label{gen_commitment}
\end{IEEEeqnarray}
\added{We depict the execution of the proposed methodology in Fig.~\ref{fig:flowchart}. Observe that the first market pass utilizes the net load scenarios, the full network model, as well as the DG cost models and operational constraints to determine the optimal DG schedules. Subsequently, the optimal DG schedules are passed onto the second pass, which enforces that the DGs committed in the first market pass remain committed in the optimal DAMC solution through~\eqref{gen_commitment}. Note that prioritizing these DGs (which are known to be able to meet the modeled net load scenarios without violating transmission constraints) for the DAM energy and FRP awards does not mean that the procured FRPs will be delivered in the face of any ramping need. However, this approach ensures that the DGs that are known in advance that will lead to stranded FRPs are not prioritized for FRP awards. It is worth emphasizing that this is precisely the target of CAISO's deployment constraints, through which CAISO aims to avoid awarding FRPs to DGs that cannot, ``ex-ante, reasonably be expected to deliver the awarded product if modeled uncertainty materializes''~\cite{caiso_frp_ferc}. Ultimately, the second pass considers the DG costs and DG operational constraints alongside the bid-in net demand and the full network model to determine the day-ahead energy and FRP awards, in addition to the LMPs and the up-FRP and down-FRP prices.}\par
We refer to our FRP procurement methodology laid out in this section as the st-FRP methodology. 
In the next section, we illustrate its relative merits on several numerical studies.

%% file: Section_III/flowchart.tex
\begin{figure*}  
\centering  
\begin{tikzpicture}[node distance=1.0cm, every node/.style={font=\footnotesize}]  
\tikzstyle{input}=[fill=COLOR_INPUT, minimum width=2.2cm, minimum height=1.2cm, align=center, line width=0pt]  
\tikzstyle{process}=[regular polygon, regular polygon sides=6, fill=COLOR_PROCESS, minimum width=1cm, minimum height=1cm, align=center, line width=0pt]  
\tikzstyle{output}=[fill=COLOR_OUTPUT, minimum width=2.5cm, minimum height=1.7cm, align=center, ellipse, line width=0pt]  
\tikzstyle{exchange}=[fill=COLOR_EXCHANGE, minimum width=1.5cm, minimum height=1.5cm, align=center, rounded corners, line width=0pt]  

\node[input] (netload) {net load \\ scenarios};  
\node[process, below=1.1cm of netload] (firstmarket) {First pass:\\ stochastic\\unit\\commitment\\model};  
\node[input, right=2.2cm of netload] (threepart) {DG production\\ and startup \\ cost models \& \\ DG operational \\ constraints};  
\node[exchange, right=1.9cm of firstmarket] (optimalschedules) {optimal schedules \\ of committed \\ generators};  
\node[input, below=1.5cm of optimalschedules] (fullnetwork) {full network \\ model};  
\node[process, right=6.5cm of firstmarket] (secondpass) {Second pass:\\day-ahead\\ market\\clearing\\ model};  
\node[input, right=2.7cm of fullnetwork] (demandbids) {demand bids};  
\node[output, right=8.5cm of threepart,  yshift=0cm] (DAMenergy) {DAM energy \\ awards};  
\node[output, below=0.22cm of DAMenergy] (LMPs) {LMPs};  
\node[output, below=0.22cm of LMPs] (DAMFRP) {DAM FRP \\ awards};  
\node[output, below=0.22cm of DAMFRP] (marketclearing) {MCPCs for\\up-FRP and\\down-FRP};
\coordinate[left=1.2cm of marketclearing] (dummy_1);  
\coordinate[left=1.2cm of DAMFRP] (dummy_2);  
\coordinate[left=1.2cm of LMPs] (dummy_3);  
\coordinate[left=1.2cm of DAMenergy] (dummy_4);  
\coordinate[right=0.5cm of secondpass] (dummy_5);  
\draw[->, line width=0.7pt, color=black!40] (secondpass) -- (dummy_5) -| (dummy_1) -- (marketclearing); 
\draw[->, line width=0.7pt, color=black!40] (secondpass) -- (dummy_5) -| (dummy_2) -- (DAMFRP); 
\draw[->, line width=0.7pt, color=black!40] (secondpass) -- (dummy_5) -|  (dummy_3) -- (LMPs); 
\draw[->, line width=0.7pt, color=black!40] (secondpass) -- (dummy_5) -|  (dummy_4) -- (DAMenergy); 
\draw[->, >=stealth, line width=0.7pt, color=black!40] (netload) -- (firstmarket);  
\draw[->, >=stealth, line width=0.7pt, color=black!40] (threepart) -- (firstmarket);  
\draw[->, >=stealth, line width=0.7pt, color=black!40] (threepart) -| (secondpass);  
\draw[->, >=stealth, line width=0.7pt, color=black!40] (fullnetwork) -| (firstmarket);  
\draw[->, >=stealth, line width=0.7pt, color=black!40] (fullnetwork) -- (secondpass);  
\draw[->, >=stealth, line width=0.7pt, color=black!40] (demandbids) -- (secondpass);  
\draw[->, >=stealth, line width=0.7pt, color=black!40] (firstmarket) -- (optimalschedules);  
\draw[->, >=stealth, line width=0.7pt, color=black!40] (optimalschedules) -- (secondpass);
\coordinate[below=0.27cm of marketclearing, xshift=1.5cm] (dummy_6);
\coordinate[left=17.9cm of dummy_6] (dummy_7);
\node[below=0.2cm of dummy_7, xshift=0.55cm] (legend) {\textbf{Legend:}};
\draw[line width=0.4pt, color=black!70] (dummy_7) -- (dummy_6);
\begin{scope}[shift={(0cm,-8cm)},every node/.append style={font=\footnotesize}]
    \node[input, minimum width=1.0cm, xshift=-0.9cm, minimum height=0.75cm] (legendInput) {};
    \node[right=0.2cm of legendInput] (legendInputLabel) {input};
    \node[process, minimum width=0.75cm, minimum height=0.75cm, right=1.4cm of legendInputLabel] (legendProcess) {};
    \node[right=0.2cm of legendProcess] (legendProcessLabel) {market pass};
     \node[exchange, minimum width=1.0cm, minimum height=0.75cm, right=1.4cm of legendProcessLabel] (legendExchange) {};
    \node[right=0.2cm of legendExchange, yshift=0.2cm] (legendExchangeLabel) {exchanged information between};
    \node[right=0.2cm of legendExchange, yshift=-0.2cm] (legendExchangeLabel2) {market passes};
    \node[output, minimum width=1.0cm, minimum height=0.75cm, right=4.9cm of legendExchange] (legendOutput) {};
    \node[right=0.2cm of legendOutput] (legendOutputLabel) {binding market outcome};
\end{scope}
\end{tikzpicture}  
\caption{\color{change_color}{Flowchart depicting the execution of the proposed methodology. Observe that the optimal schedules of the DGs committed by the first market pass are passed onto the second market pass. Subsequently, the second pass utilizes the passed information to prioritize the committed DGs for the DAM awards by enforcing that the DGs committed by the first pass remain committed in the optimal DAMC solution. Finally, the second pass issues financially binding DAM awards and outputs the LMPs and the market-clearing price for capacity (MCPC) for the up-FRP and up-FRP products.}}  
\label{fig:flowchart}  
\end{figure*}
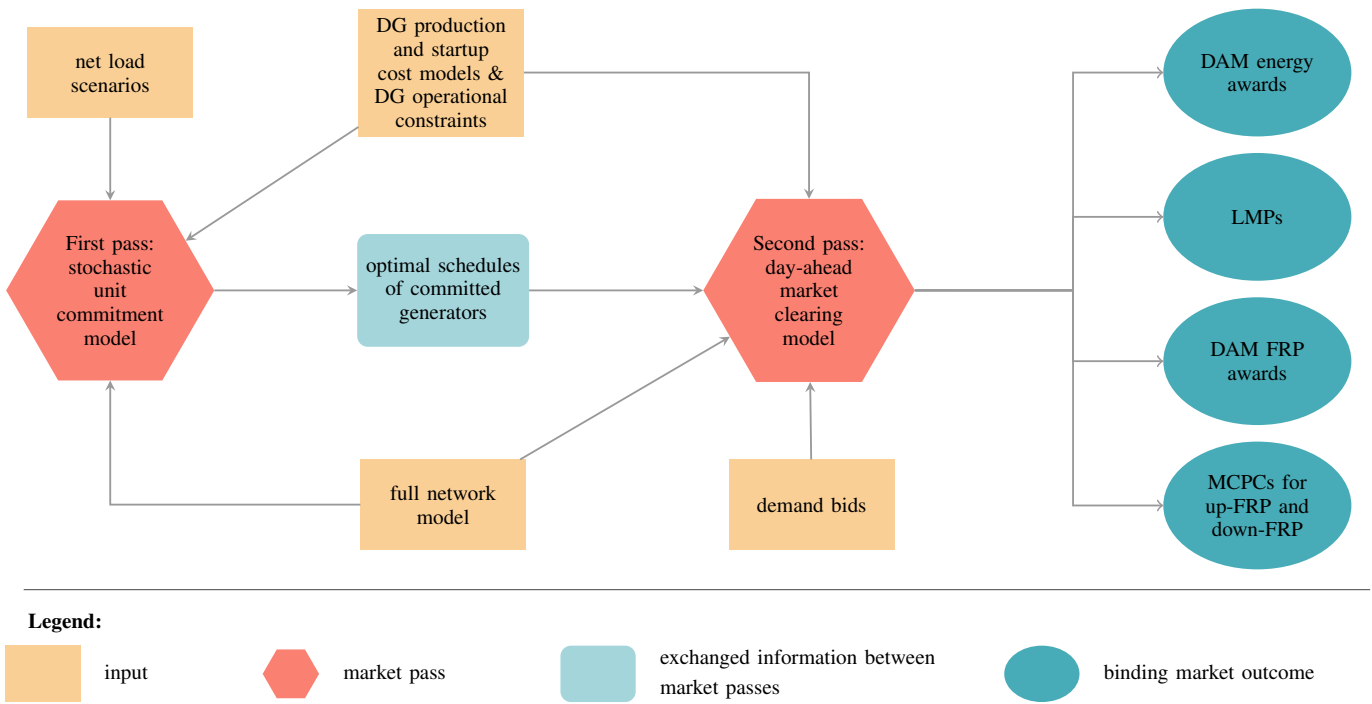

%% file: Section_IV/4intro.tex
To demonstrate the application and effectiveness of the proposed st-FRP methodology in a real-world setting, we carried out several numerical studies. 
\added{In the remainder of this section, we discuss the results on four case studies. The input data and source code of all numerical studies are provided in the online companion \cite{comp}.}

%% file: Section_IV/4a.tex
We conduct our experiments using real-world data collected in the NYISO and CAISO grids. The load data of Case Study I are constructed using the historical load values of the 11 load zones of the NYISO grid observed between April 1, 2019 and April 30, 2019~\cite{nyiso_data}. The generator and system data of Case Study I are taken from the IEEE 14-bus system, which has 11 load buses. We construct the net load data of these 11 load buses by matching each NYISO load zone to a corresponding load bus in the IEEE 14-bus test system in decreasing order of average load values. Specifically, the NYISO load zone with the highest average load is assigned to the load bus with the highest average load in the IEEE 14-bus test system, and so forth. \par
In Case Study II, we seek to assess the performance of the proposed framework using a more realistic test system and on an alternative dataset in order to simulate the ramping needs that may materialize in different months and systems. To that end, in Case Study II, we use the net load levels observed in the CAISO grid throughout December 2019~\cite{caiso_data}. The DG data of Case Study II are taken from the IEEE 300-bus system, which contains 69 DGs with an aggregate total capacity of 32,501.74 MW, thus allowing for a sufficiently realistic representation of CAISO system needs and capabilities for our purposes.\par
For performance assessments, we \added{initially} define two benchmark methods, against which we compare the performance of the proposed st-FRP methodology. One of the benchmarks we define, termed nf-FRP (nf as in \textit{not fixed}), applies the two market passes of st-FRP and sets identical FRP requirements as the proposed st-FRP method. However, nf-FRP does not fix the commitment status of the DGs through~\eqref{gen_commitment}, thus forgoing st-FRP’s prioritization scheme for FRP awards. The other benchmark method, 95-FRP, serves to reproduce the FRP procurement scheme currently used in U.S. ISOs. The 95-FRP method procures FRPs based on the 95\% confidence interval of net load, without applying the first market pass laid out in Section~\ref{sec3a}. Consequently, it does not enforce the constraint~\eqref{gen_commitment} in its market-clearing model.\par
To evaluate the performance of each method, we assess how each method does out-of-sample as an actual policy. Specifically, we begin by using each benchmark method to simulate a DAM clearing and schedule energy and FRP awards. Subsequently, we generate net load values out-of-sample, which we use to simulate an RTM\footnote{The RTM clearing simulated in our experiments follows the temporal granularity and horizon of the fifteen-minute market of CAISO. The detailed description of the RTM clearing is provided in \cite{comp}.} clearing and compute the resulting total system operation cost. We further evaluate the resulting LMPs and the up-FRP and down-FRP prices, using which we calculate the energy and FRP award settlements. When computing the payments for each DG, we compare the total cost incurred with the total payment received through its energy and FRP awards, and we schedule make-whole payments to DGs that do not otherwise break even. Finally, we use the total system operation costs, shed load amounts, as well as the FRP and energy payments computed out-of-sample to evaluate the performance of each method.\par
In our performance assessments, we assume a multivariate normal distribution for net load forecast errors, with a mean of zero and a standard deviation of three percent of the net load forecast. As modeling uncertainty is not the core focus of our work, we make the simplifying assumption that the forecast errors at different nodes and time periods are uncorrelated. We use historical net load values observed in the NYISO and CAISO grids, together with the random forecast errors, to construct net load scenarios, which we use to solve the first market pass of the st-FRP and nf-FRP methods. Finally, to simulate RTM clearing for each day in each case study, we generate an out-of-sample net load scenario using historical net load levels and a random sample of forecast errors. \par
\added{In light of the results observed from the first two case studies, we set up two additional case studies, Case Study III and Case Study IV, wherein we further assess the performance of the investigated methods. In both of these case studies, we use the DG data of the IEEE 300-bus test system and the net load values of the CAISO grid from October 1, 2019 to October 31, 2019. In addition to the nf-FRP and 95-FRP benchmark methods used in Case Studies I and II, we define two other benchmark methods in Case Studies III and IV. These additional benchmark methods, termed 90-FRP and 99-FRP, set FRP requirements based on the 90\% confidence interval and the 99\% confidence interval of net load, respectively. By introducing additional rule-based methods with varying degrees of conservatism, we seek to investigate the trade-off between cost and reliability inherent in these methods, and we analyze how the proposed method compares with them under different conditions. In these case studies, we further assess the performance of the investigated methods under autocorrelated net load forecast errors. In addition, we evaluate the out-of-sample performance, as well as the computational performance, of the methods under different scenario numbers.}\par
We conduct the experiments on a 64 GB-RAM computer containing an Apple M1 Max chip with 10-core CPU. We model all UC problems using the \verb|UnitCommitment.jl| package~\cite{UCjl}, and we solve all UC problems under Julia 1.6.1 with Gurobi 9.5.0 as the solver. 

%% file: Section_IV/4b_cs1.tex
\input{Section_IV/cs1_results}
\input{Section_IV/cs1_commitment_results}
We begin by evaluating the total system operation cost and total shed load levels delivered by each benchmark method over the entire dataset, which are laid out in Table~\ref{cs1_res}. We observe from Table~\ref{cs1_res} that the proposed st-FRP method not only yields the lowest total system operation cost among all benchmark methods, but it also completely eliminates the need for load shedding. This result shows the proposed method’s capability to schedule energy and FRP awards in a way that can efficiently meet net load ramps, while maintaining system reliability. In contrast, despite setting identical FRP requirements as st-FRP, the nf-FRP method leads to the highest total cost among all methods and requires involuntary load shedding. The higher cost and shed load levels under nf-FRP compared to st-FRP brings out the merits of using the optimal SUC solution for prioritizing the DAM awards. Finally, the 95-FRP method results in a higher total system operation cost compared to st-FRP, and it also requires load shedding, which makes evident the drawbacks of resorting to a one-size-fits-all policy to set FRP requirements. The shed load results under nf-FRP and 95-FRP signify that scheduling FRP awards under these methods are likely to require OOM actions by SOs in order to ensure system reliability. 
\par
We next turn to the DG payments under each method. Table~\ref{cs1_res} shows that the share of the FRP payments is maximized under st-FRP, highlighting the ability of the st-FRP method to explicitly acknowledge DGs for the ramping capability they provide to the system. Tellingly, the FRP payments ebb under 95-FRP, attaining values markedly lower than the other methods. These results echo the empirical observations from U.S. ISOs noted earlier in Section~\ref{sec2c}, concerning the fact that FRP procurement methods used in practice frequently lead to zero FRP prices, even during periods with dire ramping needs, thus failing to accurately reward DGs for their flexibility. The 95-FRP method further necessitates the highest make-whole payments among all methods, signifying that DGs recover a smaller portion of their costs with the energy and FRP awards. In contrast, the st-FRP method requires the lowest make-whole payments, showing the proposed method's ability to bring about more desirable cost recovery terms compared to the status quo 95-FRP method.\par
\begin{figure}[ht]
\vspace{-0.2cm}
\includegraphics[width=\linewidth]{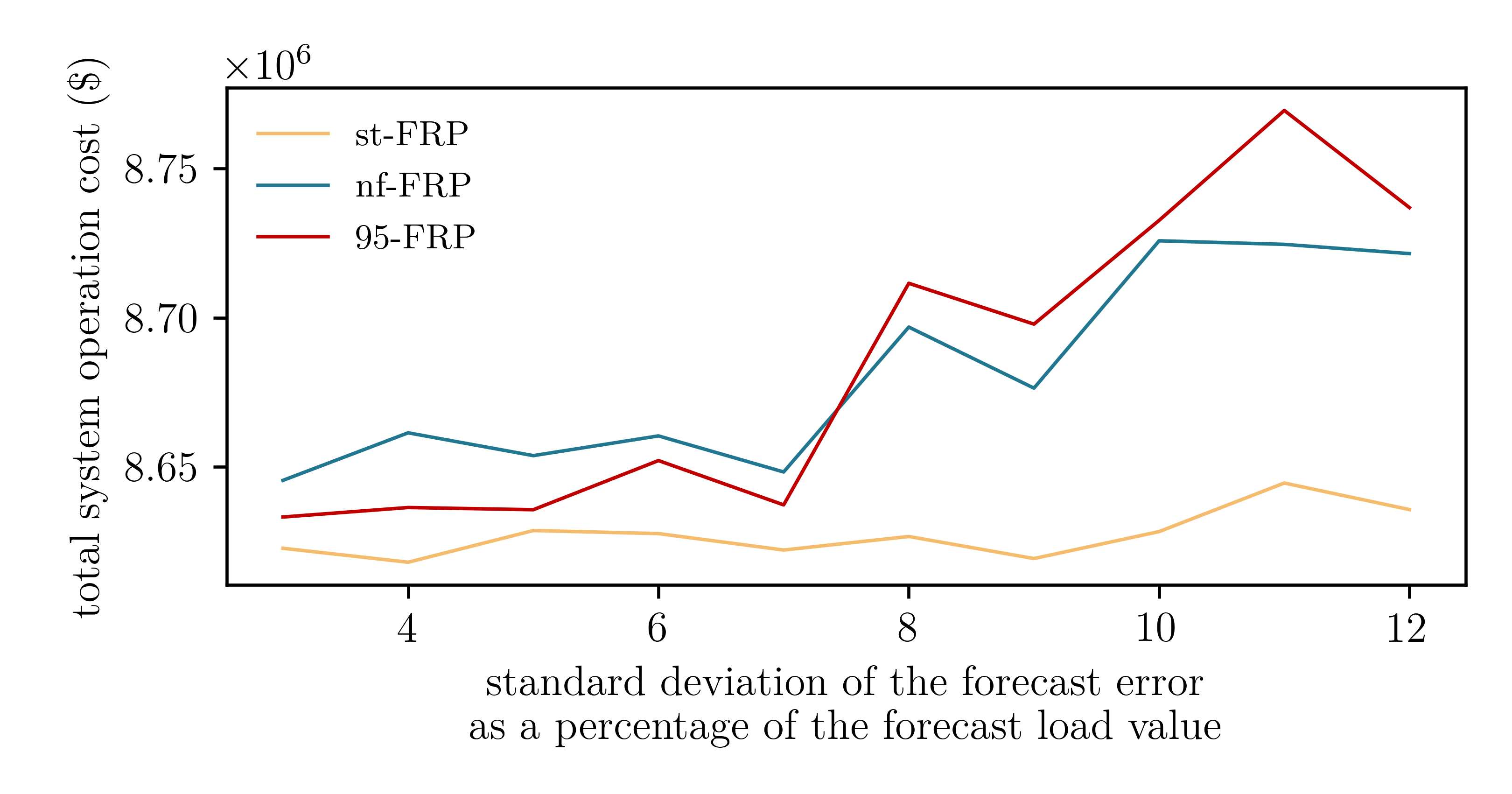}
\vspace{-0.6cm}
\caption{Total system operation costs as a function of the standard deviation level of the random forecast error used in constructing the out-of-sample scenarios.}
\vspace{-0.2cm}
\label{cs1_oos_cost}
\end{figure}
\begin{figure}[ht]
\vspace{-0.2cm}
\includegraphics[width=\linewidth]{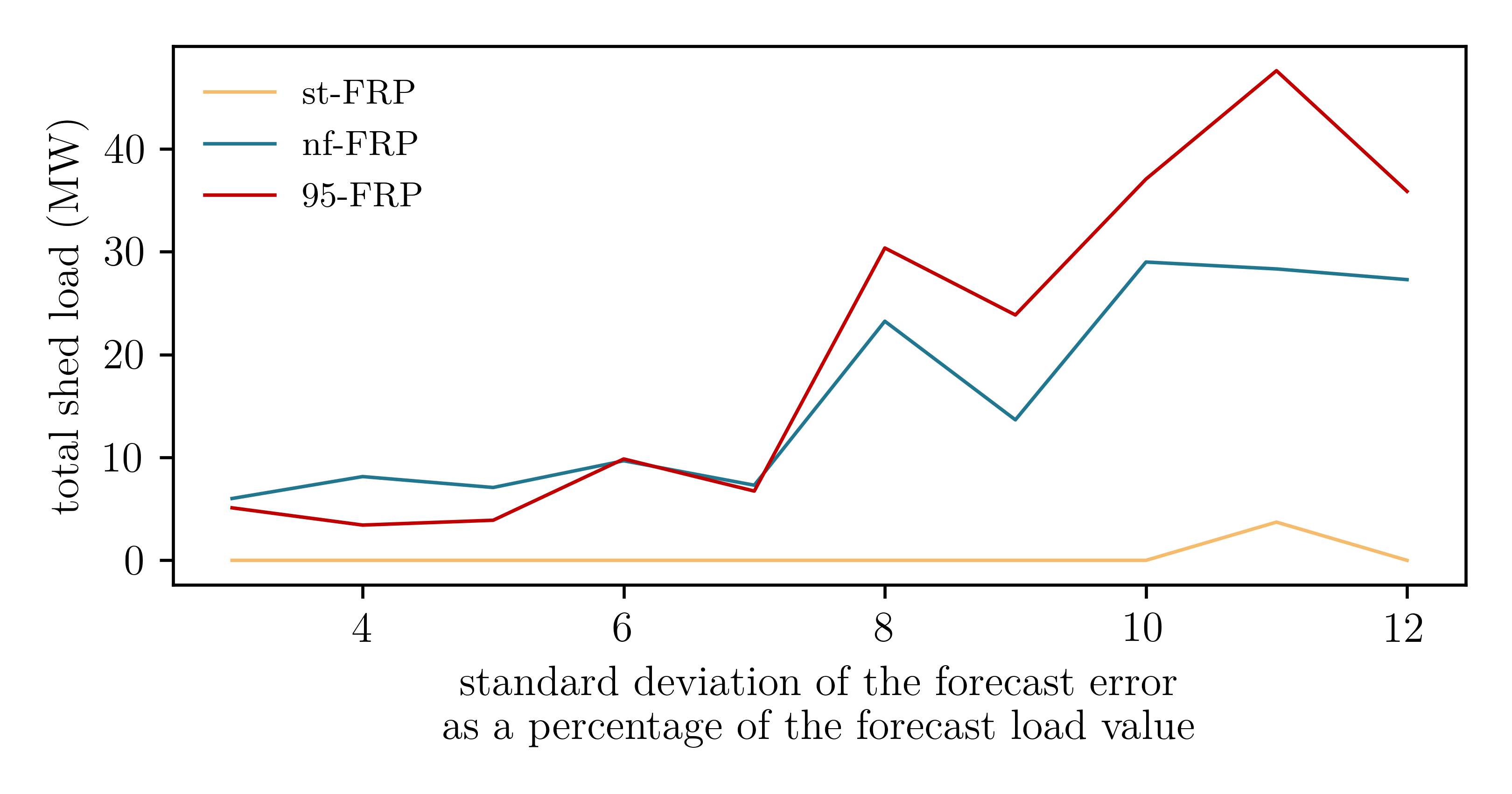}
\vspace{-0.6cm}
\caption{Total shed load levels as a function of the standard deviation level of the random forecast error used in constructing the out-of-sample scenarios.}
\vspace{-0.2cm}
\label{cs1_oos_curt}
\end{figure}

Recall that the results laid out in Table~\ref{cs1_res} are obtained under the assumption that the standard deviation of the net load forecast error is three percent of the forecasted net load value. With the deepening penetration of renewables, however, the uncertainty in net load \cite{cso_res, cso_boosting} is expected to further increase, which calls for FRP procurement methods that remain effective in the face of increased uncertainty. In this light, we assess how each method does out-of-sample under increasing degrees of net load uncertainty. In so doing, we solve the SUC problem and schedule the DAM awards based on the assumption that the standard deviation of the net load forecast error is three percent of the forecasted net load value, but we generate the out-of-sample scenarios using higher standard deviation levels. Specifically, we repeat the computation of the out-of-sample results by varying the standard deviation of the forecast error, from 3\% to 12\% of the forecasted net load value in 1\% increments. We compute the total system operation cost and shed load values over the entire dataset for different standard deviation levels and plot the results in Fig.~\ref{cs1_oos_cost} and Fig.~\ref{cs1_oos_curt}, respectively.\par
We observe from Fig.~\ref{cs1_oos_cost} that under the 95-FRP method, the total system operation costs pick up significantly as the standard deviation of the forecast error increases. In a similar vein but to a slightly lesser extent, the nf-FRP method delivers higher costs as the standard deviation of the forecast error increases. In contrast to these benchmark methods, the proposed st-FRP method remains resilient against uncertainty, with the total system operation cost trending at similar levels despite facing out-of-sample scenarios that are constructed with markedly high standard deviation levels. What’s more, Fig.~\ref{cs1_oos_curt} shows that the st-FRP method can completely obviate the need for involuntary load shedding under almost all standard deviation levels. This is in stark contrast to the curtailment results under the two benchmark methods, which sharply climb as the standard deviation attains higher values. These findings show that the st-FRP method, when compared to other benchmark methods, can more effectively and economically meet steep ramping requirements, even when the uncertainty is higher than originally accounted for.\par
\begin{figure}[ht]
\vspace{-0.2cm}
\includegraphics[width=\linewidth]{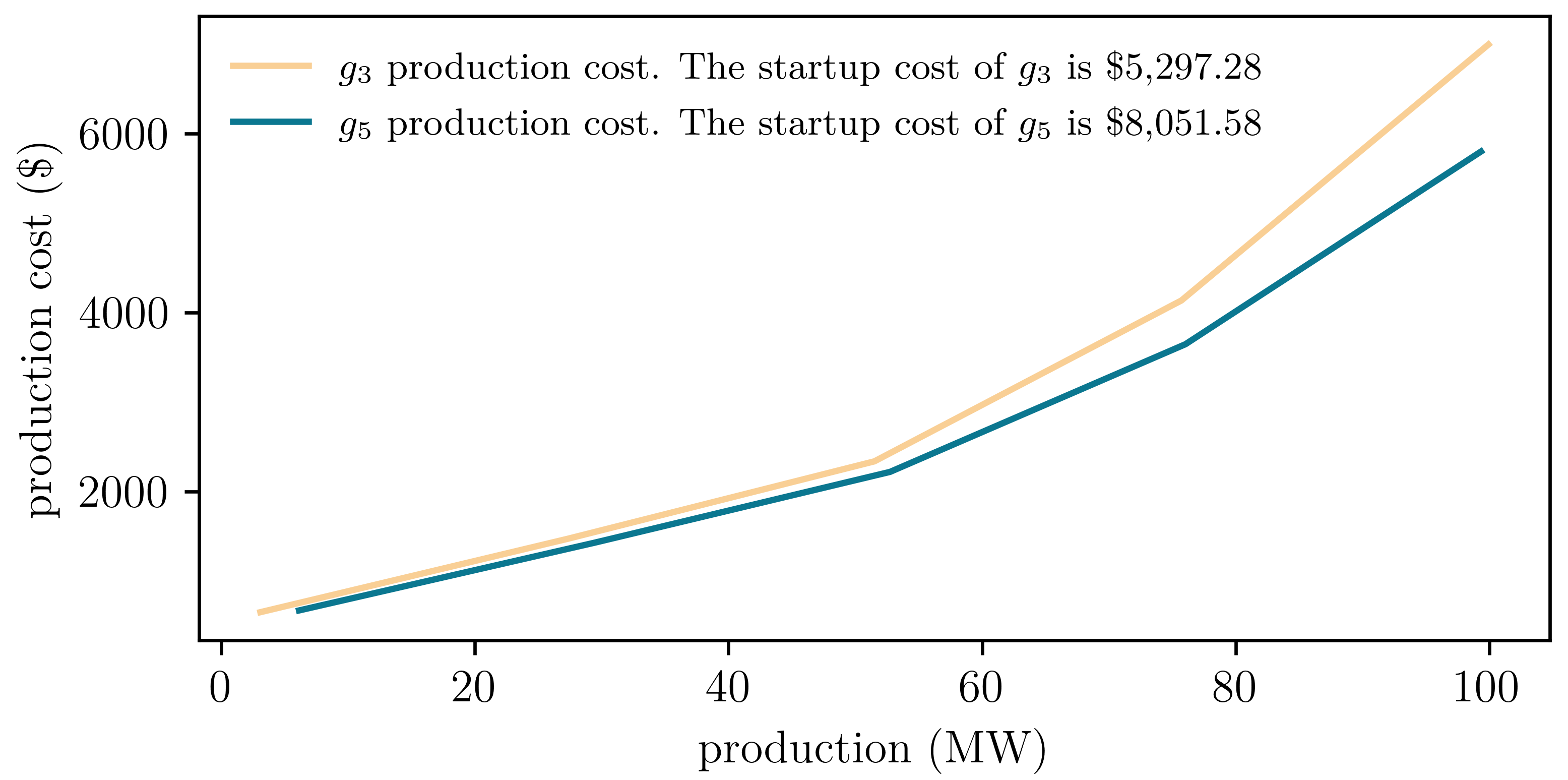}
\vspace{-0.6cm}
\caption{Production and startup costs of the generators $g_3$ and $g_5$ in the IEEE 14-bus test system used in Case Study I experiments.}
\vspace{-0.2cm}
\label{g3_g5_costs}
\end{figure}

To gain deeper insights into the observed performances, our next step is to concentrate on a specific day in the dataset and analyze the DAM awards scheduled for that day under each method. In Table~\ref{commitment}, we provide the commitment instructions received by DGs $g_3$ and $g_5$ under the st-FRP and nf-FRP methods on April 28, 2019. We observe from Table~\ref{commitment} that the st-FRP method sends commitment instructions to $g_5$ for hours 7 to 22, whereas the nf-FRP method schedules DAM awards to $g_3$ during these hours. In Fig.~\ref{g3_g5_costs}, we provide the production and start-up costs for these DGs. Fig.~\ref{g3_g5_costs} shows that while $g_3$ has a lower startup cost compared to $g_5$, its production cost is higher than that of $g_5$ for all dispatch levels. As demonstrated in Section~\ref{sec3c}, the nf-FRP method fails to acknowledge the production costs of DGs in awarding FRPs, thus sending commitment instructions to $g_3$ on the basis of its lower startup cost. Such a schedule under nf-FRP, however, yields in turn an overall higher cost when the FRPs are called into action, while also necessitating load shedding, indicating that the procured FRPs fail to meet the eventuating net load.\par
Note that while the results shown in Table~\ref{commitment} highlights one specific example, our online companion \cite{comp} provides the out-of-sample results for a wide range of experiments, revealing several other similar cases.

%% file: Section_IV/cs1_results.tex
\begin{table*}
\footnotesize
\centering
\caption{Case Study I out-of-sample results}
\renewcommand{\arraystretch}{1.5} 
\begin{tabular}{c || c | c | c | c}
\hline \hline
benchmark method &  total system operation cost (\$) &  shed load (MWh) &  FRP payment (\$) &  make-whole payment (\$) \\ \hline \hline
st-FRP &                       8,622,729.21 &              0.00 &        107,257.59 &                    0.00 \\
nf-FRP &                       8,645,499.14 &              6.01 &         46,249.53 &                 2,921.45 \\
95-FRP &                       8,633,209.55 &              5.10 &         22,521.14 &                 8,230.94 \\ \hline \hline
\end{tabular}
\label{cs1_res}
\end{table*}

%% file: Section_IV/cs1_commitment_results.tex
\begin{table*}
\footnotesize
\caption{Commitment instructions received by generators $g_3$ and $g_5$ under st-FRP and nf-FRP.}
\renewcommand{\arraystretch}{1.5} 
\begin{tabular}{l | l || rrrrrrrrrrrrrrrrrrrrrrrr}
\hline \hline 
\multicolumn{2}{c||}{}    & \multicolumn{24}{c}{hour} \\ \hline 
method & DG       &   1 &   2 &   3 &   4 &   5 &   6 &   7 &   8 &   9 &  10 &  11 &  12 &  13 &  14 &  15 &  16 &  17 &  18 &  19 &  20 &  21 &  22 &  23 &  24 \\ \hline \hline 
st-FRP & $g_3$ &   0 &   0 &   0 &   0 &   0 &   0 &   0 &   0 &   0 &   0 &   0 &   0 &   0 &   0 &   0 &   0 &   0 &   0 &   0 &   0 &   0 &   0 &   0 &   0 \\
nf-FRP & $g_3$ &   0 &   0 &   0 &   0 &   0 &   0 &   1 &   1 &   1 &   1 &   1 &   1 &   1 &   1 &   1 &   1 &   1 &   1 &   1 &   1 &   1 &   1 &   0 &   0 \\ \hline
st-FRP & $g_5$ &   0 &   0 &   0 &   0 &   0 &   0 &   1 &   1 &   1 &   1 &   1 &   1 &   1 &   1 &   1 &   1 &   1 &   1 &   1 &   1 &   1 &   1 &   1 &   1 \\
nf-FRP & $g_5$ &   0 &   0 &   0 &   0 &   0 &   0 &   0 &   0 &   0 &   0 &   0 &   0 &   0 &   0 &   0 &   0 &   0 &   0 &   0 &   0 &   0 &   0 &   0 &   0 \\ \hline \hline 
\end{tabular}
\label{commitment}
\vspace{-0.2cm}
\end{table*}

%% file: Section_IV/4b_cs2.tex
\input{Section_IV/cs2_results}
In the previous case study, we showcased the effectiveness of our methodology by applying it to the IEEE 14-bus test system and utilizing measurements from the NYISO grid. This time, our objective is to evaluate if the advantages of our methodology persist when tested with the DG data of the IEEE 300-bus test system. For this assessment, we use measurements from the CAISO grid, which features a deeper penetration of renewables and a well-established need for flexible capacity. We evaluate the out-of-sample performance of all benchmark methods using data collected throughout June 2019 and provide the results in Table~\ref{cs2_res}.\par
We observe from Table~\ref{cs2_res} that the proposed st-FRP method not only delivers the lowest total system operation cost, but also consistently meets net load in all periods, eliminating the need for load shedding. While the nf-FRP method too prevents load shedding, it does so by incurring a total system operation cost that is 2.743\% higher compared to st-FRP. Given that the only step distinguishing st-FRP from nf-FRP is prioritizing the DGs that are committed in the first market pass for receiving DAM awards, this result drives home the merits of utilizing the optimal SUC solution in scheduling the DAM awards. Finally, we remark that the 95-FRP method is the only benchmark method necessitating load shedding, signifying that the DAM awards under the 95-FRP method could require OOM actions by SOs in order to ensure system reliability.\par
Next, we turn to the FRP payments and the make-whole payments scheduled to DGs under each investigated method. In examining Table~\ref{cs2_res}, we observe that the FRP payments plummet under 95-FRP, consistent with Case~Study~I results and the findings from U.S. ISOs described in Section~\ref{sec2c}. Furthermore, the 95-FRP method requires the largest make-whole payments among all methods. In contrast, DGs receive the highest FRP payments under st-FRP, highlighting the ability of the proposed method to reward DGs explicitly for the flexible capacity they provide to the system. At the same time, the st-FRP schedules require the lowest make-whole payments among all benchmark methods. These results indicate that the proposed methodology fares better at allocating awards to flexible DGs and managing DAM schedules. This, in turn, allows DGs to recoup a larger share of their costs through transparent market awards, thereby minimizing the need for substantial OOM payments.\par
\begin{figure}[ht]
\vspace{-0.2cm}
\includegraphics[width=\linewidth]{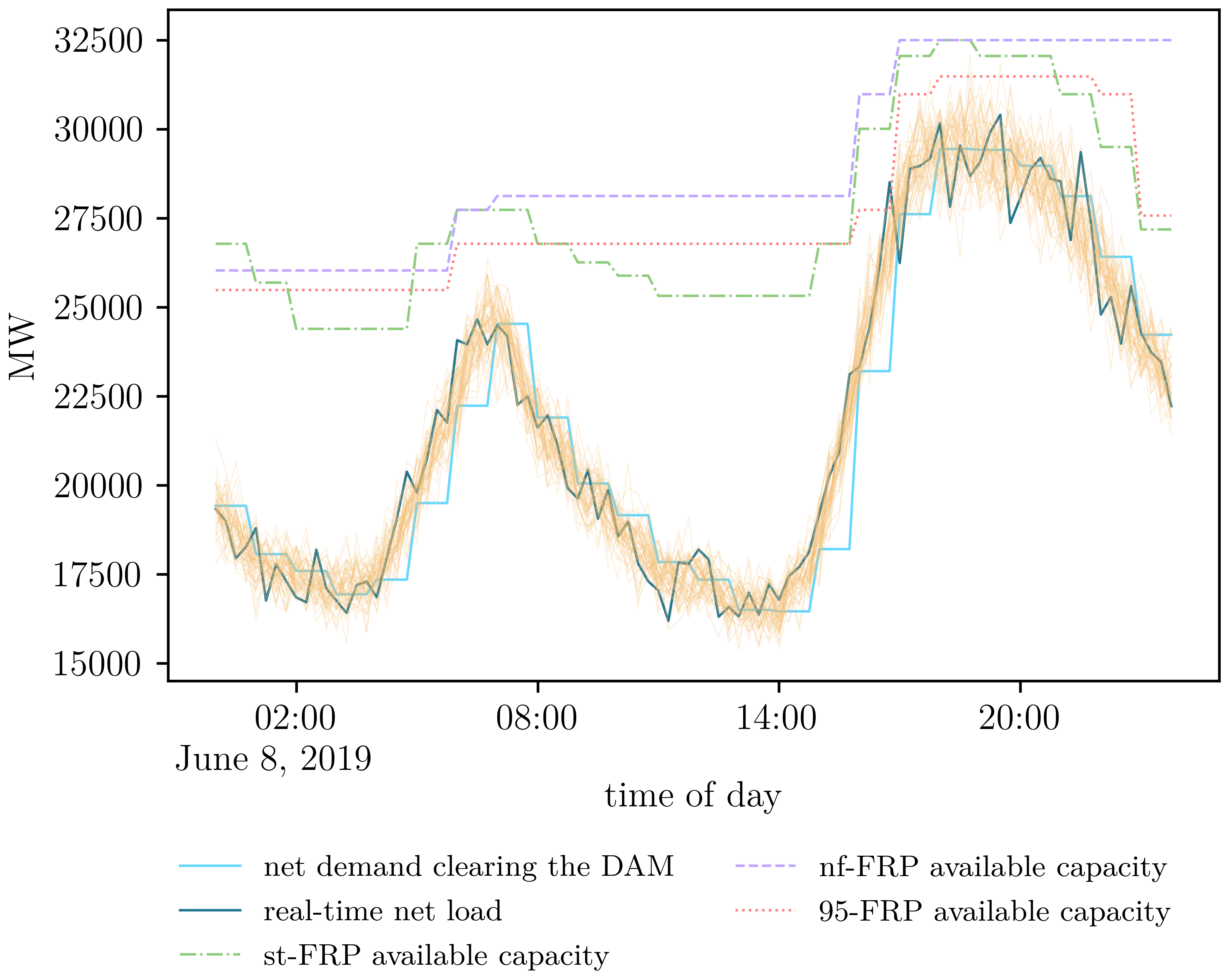}
\vspace{-0.5cm}
\caption{DAM schedules, net load scenarios, and net load levels used for out-of-sample assessments for June 8, 2019. Golden lines show the net load scenarios used in solving the first market pass under st-FRP and nf-FRP.}
\vspace{-0.2cm}
\label{cs2_plot}
\end{figure}

To unravel how the st-FRP method efficiently caters to system demand while achieving the lowest total system operation cost, we next focus on scenarios and optimal decisions for a single day in the dataset. In Fig.~\ref{cs2_plot}, we present the available capacity in the system for June 8, 2019 under all benchmark methods. Additionally, Fig.~\ref{cs2_plot} depicts the net demand that clears the DAM and the actual net load used for out-of-sample assessments. To provide additional insights, we include in Fig.~\ref{cs2_plot} the net load scenarios used in solving the first market pass of the st-FRP and nf-FRP methods.\par
Fig.~\ref{cs2_plot} illustrates the efficiency of the st-FRP method in addressing ramping needs. Achieved through the utilization of the SUC model and diverse net load scenarios, the st-FRP DAM awards ensure both economical and reliable fulfillment of net load requirements. Specifically, the DAM awards from hours ending 6 to 8 and 17 to 21 demonstrate st-FRP's ability to meet net load levels and ramping needs without excessive commitment schedules. In contrast, the 95-FRP method falls short in adequately capturing ramping needs during this time period. On top of that, the 95-FRP method results in an overcommitment of available capacity in hours ending 10 to 15, surpassing the necessary level to meet net load requirements. \added{Similar to 95-FRP, the DAM awards under the nf-FRP method lead to overcommitment in all hours after hour ending 10. This result illustrates that ignoring the commitment instructions of the first market pass could easily lead to uneconomical commitment schedules under the nf-FRP method, despite having identical FRP requirements as the st-FRP method.} The findings further bring to light the merits of employing a stochastic UC model, as seen in st-FRP, for meeting net load ramps, and accentuate the limitations of a generic rule, as seen in 95-FRP, for procuring FRPs. 

%% file: Section_IV/cs2_results.tex
\begin{table*}  
\footnotesize  
\centering  
\caption{Case Study II out-of-sample results} 
\renewcommand{\arraystretch}{1.5}   
\begin{tabular}{c || c | c | c | c}  
\hline \hline  
benchmark method &  total system operation cost (\$) &  shed load (MWh) &  FRP payment (\$) &  make-whole payment (\$) \\ \hline \hline  
st-FRP &                     \$1.25B &              0.00 &     \$322.68M &            \$12.47M \\  
nf-FRP &                     \$1.28B &              0.00 &     \$189.10M &            \$66.45M \\  
95-FRP &                     \$1.27B &            630.07 &     \$0.24M &            \$145.75M \\ \hline \hline  
\end{tabular}  
\label{cs2_res}  
\end{table*}

%% file: Section_IV/4b_cs3.tex
\input{Section_IV/cs3_results}
\added{In the previous case studies, we compared the performance of the proposed st-FRP method with that of the nf-FRP and 95-FRP methods, where the 95-FRP method was designed to replicate the approach currently used by CAISO, MISO, and SPP in setting FRP requirements. To gain further insights into how the percentiles used in setting the FRP requirements influence the out-of-sample performances, we next introduce two additional benchmark methods: 90-FRP and 99-FRP. While the 90-FRP method sets FRP requirements based on the 90\% confidence interval of net load, the more conservative 99-FRP method procures FRPs to cover the 99\% confidence interval.}\par
\begin{figure*}
\centering
    \subfloat[][\label{cs3_a}]{\includegraphics[width=.497\linewidth]{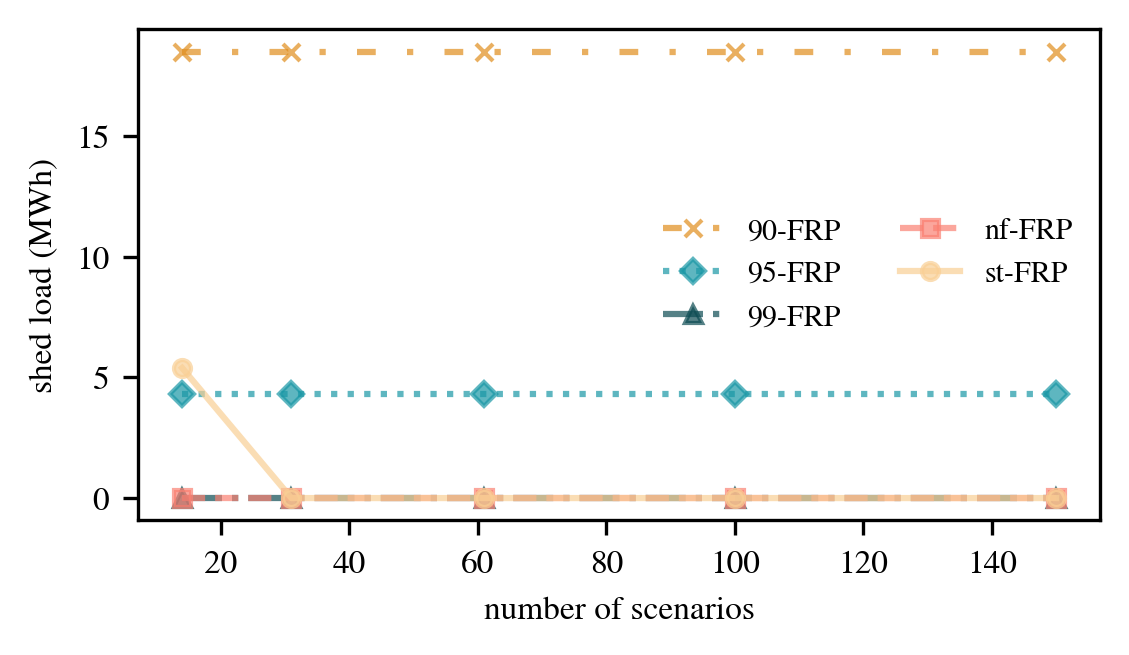}}\hspace{0.0cm}
    \subfloat[][\label{cs3_b}]{\includegraphics[width=.497\linewidth]{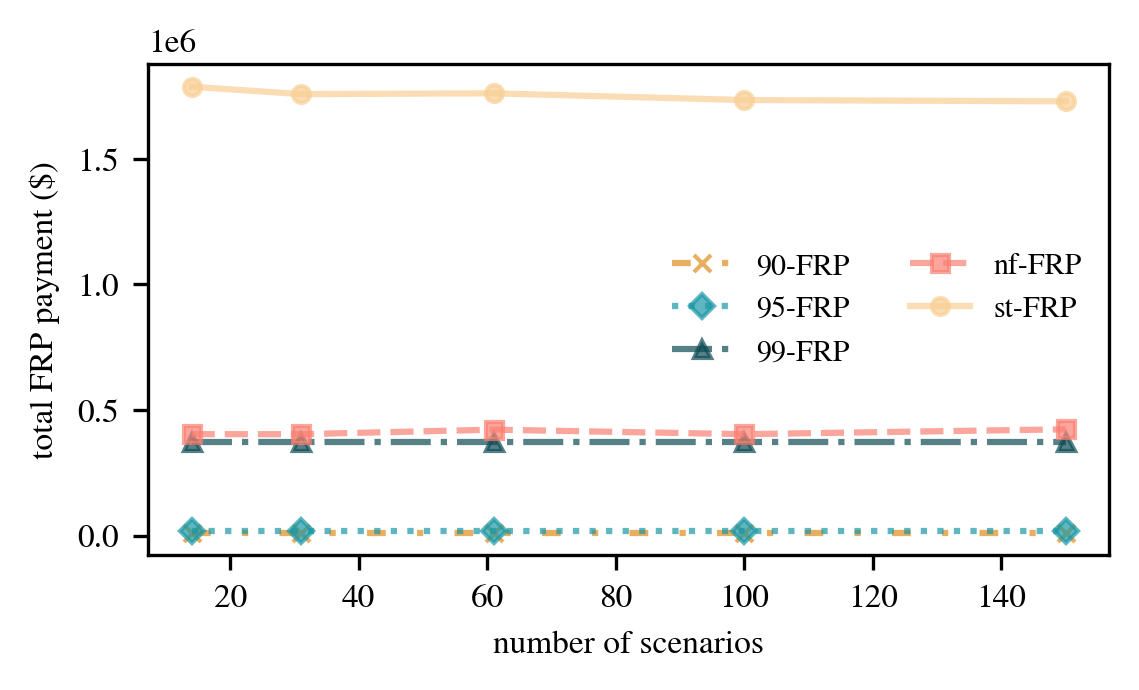}}%
    \caption{\color{change_color}{Case Study III out-of-sample results. Fig.~\ref{cs3_a} depicts the total shed laod results, and Fig.~\ref{cs3_b} the total FRP payments under each method for different scenario numbers over the entire Case~Study~III dataset. Observe that the the st-FRP method can completely eliminate the need for load shedding for all scenario numbers greater than 14. In contrast, the conservative 99-FRP method, as well as the nf-FRP method, does not require load shedding under any of the tested number of scenarios. Fig.~\ref{cs3_b} shows that while the FRP payments are maximized under the proposed st-FRP method, they plummet under the 95-FRP and the least conservative 90-FRP methods.}}
\label{cs3_figure}
\end{figure*}   
In addition to introducing two other benchmark methods, in the next two case studies, we perform additional tests to evaluate the performance of the methods under different system conditions. One additional test we conduct is to analyze how the number of scenarios modeled in the SUC formulation influences the out-of-sample performances and the computational needs of the st-FRP and nf-FRP methods. Furthermore, we investigate how different methods perform under autocorrelated net load forecast errors. Given that the net load forecast errors in adjacent time periods are autocorrelated, we check how the out-of-sample performances are influenced by the autocorrelation of forecast errors. To that end, we generate the random net load forecast errors based on an autoregressive AR(1) model under different values of the first-order autocorrelation coefficient $\rho$. Specifically, for each sub-period $k \in \mathscr{K}$, we construct the net load scenario $\xi^{\omega},\;\omega \in \Omega$ as
\begin{IEEEeqnarray}{l}
\xi^{\omega}[k] \coloneqq  \overline{\xi}[k] + \epsilon^{\omega}[k]
\end{IEEEeqnarray}
where $\overline{\xi}[k]$ denotes the expected net load in sub-period $k$ and $\epsilon^{\omega}[k]$ the random net load forecast error in sub-period $k$ and scenario $\omega \in \Omega$. Following the AR(1) model, we construct the random net load forecast errors as
\begin{IEEEeqnarray}{l}
\epsilon^{\omega}[k] \coloneqq \rho \epsilon^{\omega}[k-1] + \sqrt{1-\rho^2} \eta^{\omega}[k] 
\end{IEEEeqnarray}
where $\rho$ denotes the autocorrelation coefficient, $\epsilon^{\omega}[k-1]$ the random net load forecast error in sub-period $k-1$ and scenario $\omega \in \Omega$, $\eta^{\omega}[k]$ the random shock in period $k$ and scenario $\omega \in \Omega$, and the scaling with $\sqrt{1-\rho^2}$ serves to ensure that the variance of the random net load forecast errors stays consistent across time periods. The random shock term $\tilde{\eta}[k]$ in sub-period $k$ has the distribution
\begin{IEEEeqnarray}{l}
\tilde{\eta}[k]\, \sim \, \mathcal{N} (0, \sigma[k]) 
\end{IEEEeqnarray}
where the standard deviation term $\sigma[k]$ is taken to be three percent of the net load forecast, as described previously in Section~\ref{sec4a}. Finally, we construct the random shock term for each scenario $\eta^{\omega}[k], \omega \in \Omega$, by generating an independent and identically distributed random sample of $\tilde{\eta}[k]$ of $|\Omega|$ realizations. Note that for the first sub-period, $\rho$ is taken to be zero so the random net load forecast error for $k=1$ in each scenario $\omega \in \Omega$ is constructed using solely the random shock in each scenario $\eta^{\omega}[k], \omega \in \Omega$.
To ensure that the results obtained with different scenario numbers and autocorrelation levels can be compared on a consistent basis, in the next two case studies, we fix the value of the seed used to generate the random in-sample and out-of-sample scenarios. In Case Study III, we begin by focusing on the case $\rho = 0.0$ (i.e., not modeling the forecast errors as autocorrelated), and we investigate in Case Study IV the cases where $\rho \in \{0.2, 0.4, 0.6, 0.8\}$.\par
\added{The results in Table~\ref{table_cs3_res} show that the proposed st-FRP method can meet net load at a lower total cost compared to all benchmark methods, while completely eliminating the need for load shedding. At the same time, the st-FRP method leads to the largest FRP payments among all tested methods, and it is the only method that does not necessitate any make-whole payment. Mirroring the conclusions from the previous case studies, these results underscore that the st-FRP schedules not only reliably and economically meet net load, but also send appropriate price signals for flexibility and lead to more favorable cost recovery conditions. Among the remaining methods, the total system operation cost attains its lowest ebb under the 99-FRP method, which, similar to st-FRP, does not require any load shedding. Indeed, by seeking to cover the 99\% confidence interval of net load, the higher FRP requirements set under 99-FRP lead to higher FRP payments, as well as lower make-whole payments, compared to 95-FRP and 90-FRP. In contrast, the smaller FRP requirements set under the least conservative 90-FRP method bring about the lowest FRP payments and necessitate the highest amount of load shedding among all tested methods. Apart from the proposed st-FRP method, the nf-FRP schedules lead to the highest FRP payment, and they also eliminate the need for load shedding. However, the total system operation cost under nf-FRP is higher than all but one method, indicating how cost efficiency can suffer from not using the optimal SUC solution in prioritizing the DAM awards.}\par
\added{Recall that the st-FRP and nf-FRP results provided in Table~\ref{table_cs3_res} are obtained using 100 in-sample net load scenarios. Our next step is to analyze how the out-of-sample performances of these methods change under varying number of in-sample scenarios. To that end, we repeat the experiments under 14, 31, 61, 100, and 150 in-sample scenarios. For each case, we compute the out-of-sample performances of all methods over the entire out-of-sample dataset and plot the load shedding and FRP payment results in Fig.~\ref{cs3_figure}}.\par
\input{Section_IV/cs4_results}
\begin{figure*}    
\centering
    \subfloat[][\label{cs4_a}]{\includegraphics[width=.48\linewidth]{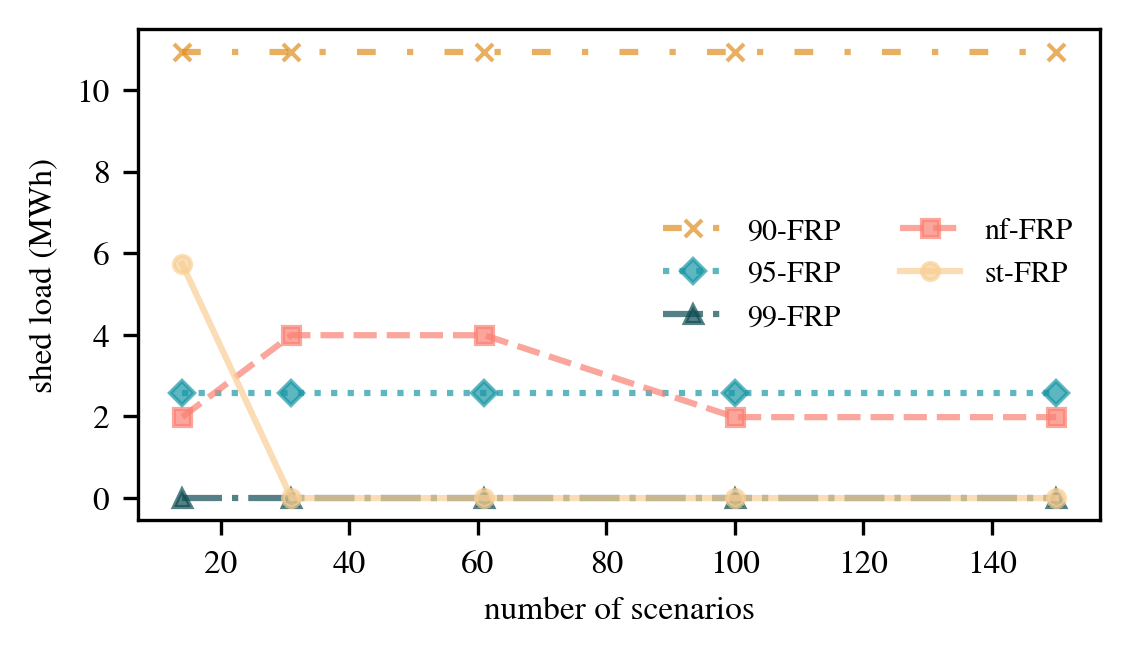}}\hspace{0.0cm}
    \subfloat[][\label{cs4_b}]{\includegraphics[width=.51\linewidth]{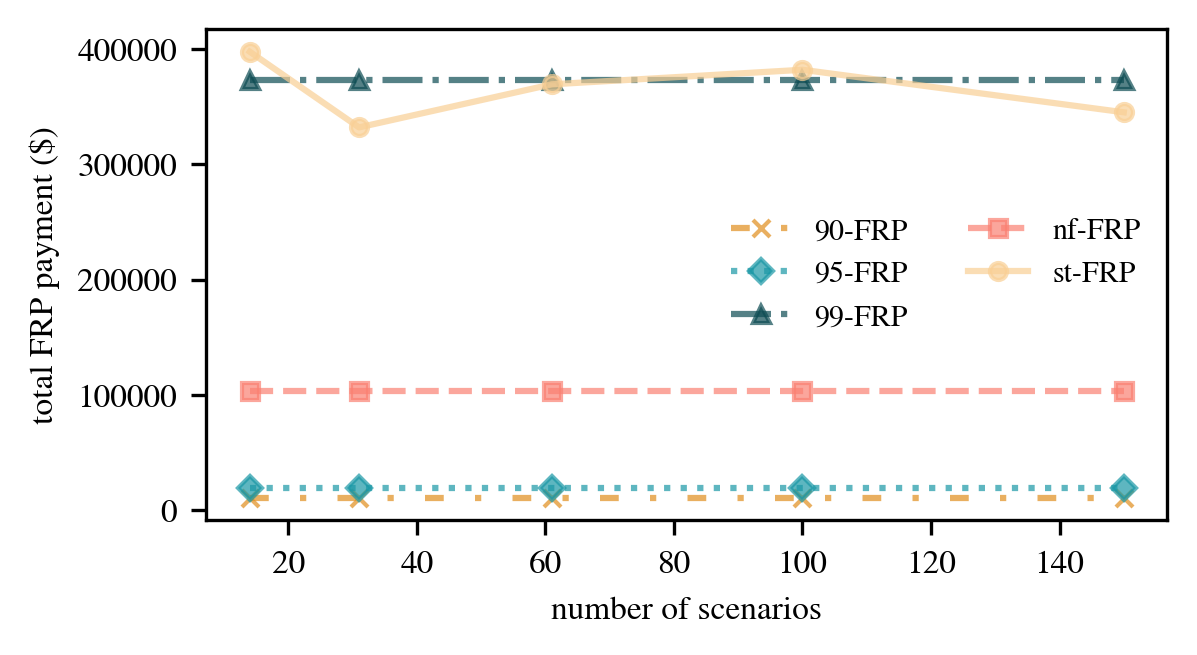}}%
    \caption{\color{change_color}{Total shed load (Fig.~\ref{cs4_a}) and total FRP payment (Fig.~\ref{cs4_b}) results evaluated out-of-sample over the entire dataset in Case Study IV. In a similar vein as the previous case studies, st-FRP not only eliminates the need for load shedding under all scenario numbers greater than 14, but also handsomely awards DGs for their flexibility. Note that while similar results are obtained for 99-FRP, st-FRP yields these favorable conditions with least total system operation cost, as laid out in Table~\ref{res_cs4_table} and Table~\ref{rho_res_cs4}.} }
\label{cs4_fig}
\end{figure*}
\added{We observe from Fig.~\ref{cs3_a} that the proposed st-FRP method can completely eliminate the need for load shedding for all tested number of scenarios greater than 14. In contrast, the nf-FRP method completely avoids load shedding under all tested scenario numbers. Fig.~\ref{cs3_b} shows that the FRP payments attain their highest levels under st-FRP for all tested scenario numbers, followed by the nf-FRP method. Note that the out-of-sample results under the 90-FRP, 95-FRP, and 99-FRP methods do not depend on the number of scenarios, as these methods do not make use of the SUC solution.}

%% file: Section_IV/cs3_results.tex
\begin{table*}  
\footnotesize
\centering  
\renewcommand{\arraystretch}{1.5}   
\caption{Case Study III Out-of-Sample Results}  
\begin{tabular}{c || c | c | c | c}  
\hline \hline  
benchmark method &  total system operation cost (\$) &  shed load (MWh) &  FRP payment (\$) &  make-whole payment (\$) \\ \hline \hline  
st-FRP & 9,077,987.51 & 0.00 & 1,734,014.13 & 0.00 \\  
nf-FRP & 9,162,577.33 & 0.00 & 403,961.54 & 11,815.93 \\  
90-FRP & 9,167,270.92 & 18.50 & 10,293.72 & 8,503.08 \\  
95-FRP & 9,110,004.02 & 4.33 & 19,379.22 & 8,841.59 \\  
99-FRP & 9,107,478.22 & 0.00 & 373,596.89 & 817.29 \\ \hline \hline  
\end{tabular}
\label{table_cs3_res}  
\end{table*}

%% file: Section_IV/cs4_results.tex
\begin{table*}
\footnotesize	  
\centering   
\renewcommand{\arraystretch}{1.5}   
\caption{Case Study IV Out-of-Sample Results} 
\begin{tabular}{c || c | c | c | c}  
\hline \hline  
benchmark method &  total system operation cost (\$) &  shed load (MWh) &  FRP payment (\$) &  make-whole payment (\$) \\ \hline \hline  
st-FRP & 9,043,155.87 & 0.00 & 382,197.33 & 0.00 \\  
nf-FRP & 9,078,520.46 & 1.98 & 102,905.30 & 3,697.96 \\  
90-FRP & 9,091,490.32 & 10.94 & 10,293.72 & 8,503.08 \\  
95-FRP & 9,063,730.66 & 2.57 & 19,379.22 & 8,917.32 \\  
99-FRP & 9,070,521.51 & 0.00 & 373,596.89 & 826.56 \\ \hline \hline  
\end{tabular}
\label{res_cs4_table}  
\end{table*}  

%% file: Section_IV/4b_cs4.tex
\added{In Case Study IV, our primary objective is to assess how the out-of-sample performances change with autocorrelated net load forecast errors. To that end, we use the exact system and net load data as Case~Study~III, with the exception that the in-sample and out-of-sample net load scenarios are generated with different values of $\rho$. We begin by discussing the results under $\rho = 0.6$, which are provided in Table~\ref{res_cs4_table}.}\par
\input{Section_IV/cs4_results_2}
\begin{figure*}    
\centering
   {\includegraphics[width=\linewidth]{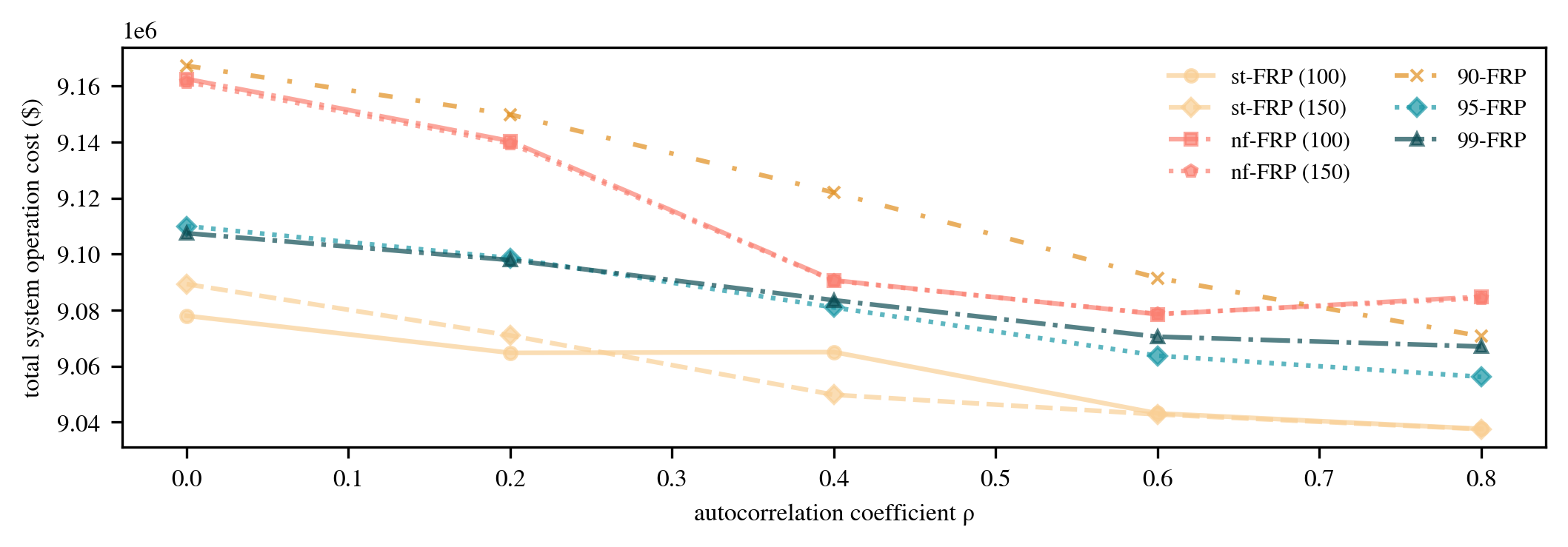}}
    \caption{\color{change_color}{Total system operation cost for different benchmark methods and scenario numbers under various values of the autocorrelation coefficient $\rho$. Observe that under almost all methods, total system operation costs decrease as $\rho$ increases. This can be attributed to less severe net load fluctuations that result from highly autocorrelated net load forecast errors, leading in turn to smaller ramping needs in the system.}}
    \label{cs4_rho_fig}
    \vspace{-0.5cm}
\end{figure*}
\added{We observe from Table~\ref{res_cs4_table} that the st-FRP method can meet net load at least total cost, without necessitating any load shedding. Furthermore, the st-FRP schedules lead to the lowest make-whole payments and highest FRP payments among all tested methods. These results reinforce the relative strengths of st-FRP identified in previous studies. Comparing the rule-based 90-FRP, 95-FRP, and 99-FRP methods, we observe that as the confidence interval considered in setting the FRP requirements increases, the shed load levels decrease while the FRP payments increase. Importantly, the 99-FRP method yields lower shed load levels but higher total system operation costs compared to the 95-FRP, making clear the trade-off between reliability and costs when it comes to setting FRP requirements through rule-based mechanisms. The st-FRP method, however, delivers the best of both worlds by bringing about lower total system operation costs compared to 95-FRP while fully eliminating load shedding as in 99-FRP. We observe from Fig.~\ref{cs4_a} that the st-FRP method avoids load shedding for all tested scenario numbers greater than 14. At the same time, st-FRP yields either the highest or the second highest FRP payments under all tested scenario numbers, as shown in Fig.~\ref{cs4_b}.}\par
\added{Comparing the results of Case~Study~III with those of Case~Study~IV, we observe that the proposed st-FRP method consistently delivers the lowest total system operation cost and shed load levels, while handsomely rewarding DGs for the flexibility they provide to the system through transparent FRP awards. In contrast, the performance of rule-based methods fluctuates under different values of $\rho$. Considering 90-FRP, 95-FRP, and 99-FRP, while 99-FRP leads to the lowest out-of-sample cost in Case~Study~III, the 95-FRP methods outperforms 99-FRP in Case~Study~IV in terms of total system operation costs. However, 95-FRP fails to avoid load shedding in both case studies and necessitates the highest make-whole payments in Case~Study~IV, demonstrating that the 95-FRP method falls short of ensuring a reliable system operation despite leading to a lower total cost. Finally, while the nf-FRP method avoids load shedding in Case~Study~III, it requires load shedding in all tested scenario numbers in Case~Study~IV. Table~\ref{res_cs4_table} shows that the nf-FRP method leads to higher total system operation costs in Case~Study~IV compared to all but one of the benchmark methods, which was also noted in Case~Study~III. These results reaffirm the benefits of using the optimal SUC solution for scheduling the DAM energy and FRP awards, as applied in the proposed st-FRP method.}
\added{In Table~\ref{rho_res_cs4}, we provide the total system operation cost obtained with varying autocorrelation levels under different benchmark methods and scenario numbers. For ease of exposition, we further plot these results for a select group of benchmark methods in Fig.~\ref{cs4_rho_fig}.}
\added{Table~\ref{rho_res_cs4} shows that under all tested values of $\rho$, the proposed st-FRP method delivers the lowest total system operation cost for all scenario numbers greater than 14. This result confirms the findings from earlier case studies, making clear that the st-FRP method remains cost effective even when the autocorrelation of net load forecast errors is modeled at varying levels. Under almost all methods, the total system operation cost reduces as the value of $\rho$ increases, controlling for the number of scenarios in the st-FRP and nf-FRP methods. This result can be attributed to the fact that the forecast errors in adjacent time periods are more likely to materialize in the same direction under increasing values of $\rho$, leading in turn to less severe net load fluctuations and ramping needs.}\par
\added{An important consideration is the influence of the number of scenarios on the out-of-sample performance of the st-FRP and nf-FRP methods. Analyzing each tested value of $\rho$, we observe that there is no clear trend in the total system operation costs of the st-FRP and nf-FRP methods under changing scenario} \added{numbers beyond 31. While increasing the number of scenarios from 14 to 31 leads to lower total system operation costs under all values of $\rho$ for the st-FRP method, no such trend is observed for scenario numbers greater than 31.}\par
\begin{figure}    
\centering
   {\includegraphics[width=\linewidth]{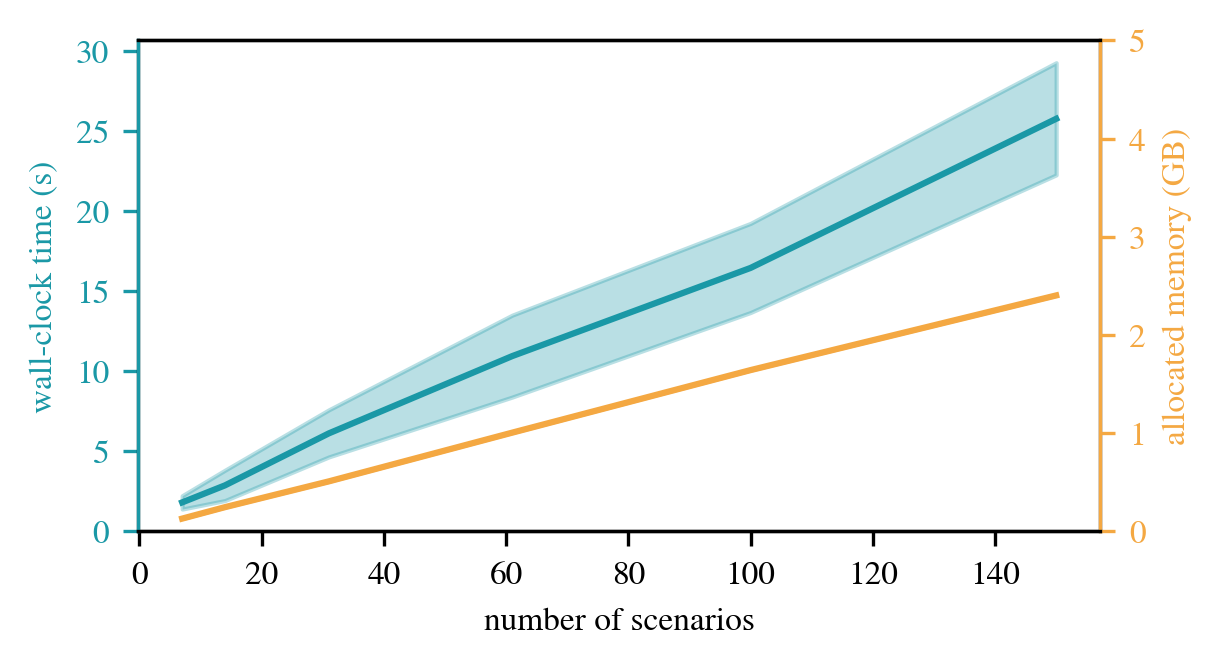}}
   \vspace{-0.5cm}
    \caption{\color{change_color}{Wall-clock time and allocated memory for solving the SUC problem under varying number of scenarios. The solid line depicts the average wall-clock time for solving the SUC problem over all days in the dataset, and the shaded region indicates one sample standard deviation around the mean.}}
    \label{solution_performance}
    \vspace{-0.5cm}
\end{figure}

\added{It is important to note that the number of scenarios required to model net load depends greatly on the dimension of the parameter and the underlying characteristic of uncertain net load, which in and of itself is nonstationary and can significantly change with the resource mix of the system in question. As such, these results can by no means generalized to other systems and experiments. However, the trade-off between cost efficiency and the number of scenarios is critical for decision-making methods that leverage scenarios, as the number of scenarios can markedly impact the solution time and the computational requirements of these methods. This plays a} \added{vital role in determining the feasibility of a decision-making method in real-world power system applications, because the} \added{market-clearing engines of system operators have stringent time limits for solving their security-constrained UC models and publishing market outcomes.} \added{In this light, we plot in {Fig.~\ref{solution_performance}} the solution time and the memory usage for solving the SUC problem as a function of the number of scenarios for $\rho=0.6$. Since we solve the extensive form of the SUC problem, we observe that the memory usage and solution time increase linearly with the number of scenarios.}\footnote{\added{The reported wall-clock times could be significantly improved by using solution techniques such as progressive hedging or the L-shaped method, which can efficiently exploit parallel computing. See \cite{suc_ph} for the application of progressive hedging to the SUC problem.}} \added{In real-world applications, solution times need to be assessed in conjunction with out-of-sample performances to determine the number of scenarios that ensure the optimal solution is obtained within the applicable time limits without significantly compromising on out-of-sample performance.}

%% file: Section_IV/cs4_results_2.tex
\begin{table*}
\footnotesize
\centering
\renewcommand{\arraystretch}{1.5}
\caption{Total system operation cost (\$) under different benchmark methods and number of scenarios for different values of $\rho$}
\begin{tabular}{c || c | c | c | c | c}
\hline \hline
benchmark method (number of scenarios) & $\rho$=0.0 & $\rho$=0.2 & $\rho$=0.4 & $\rho$=0.6 & $\rho$=0.8 \\ \hline \hline
st-FRP (14) & 9,099,659.61 & 9,064,396.53 & 9,049,633.86 & 9,059,013.79 & 9,089,737.67 \\
st-FRP (31) & 9,078,762.17 & 9,075,620.07 & 9,049,865.95 & 9,037,111.49 & 9,033,217.20 \\
st-FRP (61) & 9,081,120.15 & 9,068,460.24 & 9,049,609.11 & 9,040,609.24 & 9,033,131.66 \\
st-FRP (100) & 9,077,987.51 & 9,064,754.74 & 9,065,031.04 & 9,043,155.87 & 9,037,613.81 \\
st-FRP (150) & 9,089,299.49 & 9,070,958.20 & 9,049,778.02 & 9,042,756.66 & 9,037,491.63 \\ \hline
nf-FRP (14) & 9,162,107.29 & 9,138,690.43 & 9,092,728.06 & 9,078,831.20 & 9,079,258.43 \\
nf-FRP (31) & 9,163,036.55 & 9,140,782.72 & 9,092,642.22 & 9,088,606.61 & 9,079,130.90 \\
nf-FRP (61) & 9,161,478.08 & 9,139,821.07 & 9,091,890.77 & 9,088,858.80 & 9,084,676.13 \\
nf-FRP (100) & 9,162,577.33 & 9,140,378.60 & 9,090,664.20 & 9,078,520.46 & 9,084,803.66 \\
nf-FRP (150) & 9,161,465.53 & 9,139,517.28 & 9,090,465.69 & 9,078,668.99 & 9,084,262.49 \\ \hline
90-FRP & 9,167,270.92 & 9,149,843.82 & 9,122,014.92 & 9,091,490.32 & 9,070,658.05 \\ \hline
95-FRP & 9,110,004.02 & 9,098,603.56 & 9,080,978.60 & 9,063,730.66 & 9,056,200.04 \\ \hline
99-FRP & 9,107,478.22 & 9,097,884.09 & 9,083,553.46 & 9,070,521.51 & 9,067,026.75 \\
\hline \hline
\end{tabular}
\label{rho_res_cs4}
\end{table*}

%% file: Section_V/5.tex
In this paper, we discussed a methodology for procuring flexible ramping products (FRPs) in the day-ahead market (DAM). There are important policy implications arising from our analysis. First, we identified that employing a stochastic unit commitment (SUC) model to determine the FRP requirements could be of great help in accurately identifying and meeting sharp changes in net load. Specifically, we observed how the optimal SUC solution helps pinpoint the severe ramping needs that arise in critical hours---all while avoiding the overcommitment of generators in less volatile and uncertain periods. In contrast, the 95-FRP method, akin to methodologies used by U.S. ISOs to schedule FRP awards, can fall short of meeting precipitous net load ramps, and it occasionally introduces excess capacity into the system. Second, we found that scheduling the FRP awards based on the optimal SUC solution delivers not only a more reliable and economical system operation, but also proves conducive to favorable cost recovery profiles for generators. This is evidenced by the st-FRP method entailing lower total system operation costs and make-whole payments vis-à-vis nf-FRP in both case studies. Despite setting identical FRP requirements, the st-FRP method's prioritization scheme more favorably rewards generator flexibility and leads to the delivery of a greater portion of the scheduled FRP awards compared to nf-FRP. Finally, and importantly, we noted that the status quo 95-FRP method oftentimes leads to tight supply conditions, causing extreme price spikes in the real-time market (RTM). In contrast, the proposed st-FRP method can avoid such conditions in a wide array of instances, contributing to less volatile LMPs in the RTM. \par
In our future work, we seek to integrate risk measures \cite{risk_opt} into the proposed methodology so as to explicitly quantify the risk of not meeting net load ramps. Another avenue for future research is to incorporate other reserve products into our approach and investigate the interplay between operating reserves and FRPs in scheduling the DAM awards.

%% file: Appendices/nomenclature.tex
\added{In this appendix, we provide the nomenclature used in the publication.}
\subsection*{\added{Sets,Indices}}
{
\color{change_color}{
\begin{xtabular}[h]{ll}
$\mathscr{H}, h$  &\hspace{0.38cm} set, index of hourly periods of a day\\
$\mathscr{K}, k$  &\hspace{0.38cm} set, index of intra-hourly sub-periods of a day\\
$\mathscr{N}, n$  & \hspace{0.38cm} set, index of nodes\\
$\mathscr{L}, \ell$  & \hspace{0.38cm} set, index of lines\\
$\mathscr{G}, g$  & \hspace{0.38cm} set, index of DGs\\
$\mathscr{S}_g, s$ & \hspace{0.38cm} set, index of piecewise cost intervals for DG $g$\\
$\Omega, \omega$ & \hspace{0.38cm} set, index of net load scenarios
\end{xtabular}}}
\subsection*{\added{Variables}}\label{2}
{
\color{change_color}{
\begin{xtabular}[h]{ll}
\multicolumn{2}{l}{\textit{Note: variables are indexed by the hour $(h)$ or the}}\\
\multicolumn{2}{l}{\textit{intra-hourly sub-period $[k]$ in different formulations}}\\
\multicolumn{2}{l}{\textit{laid out in the publication.}}\\
$u_{g}$ & \hspace{0.55cm} $\in \{0,1\}$, commitment status of DG $g$\\
$v_{g}$ & \hspace{0.55cm} $\in \{0,1\}$, startup status of DG $g$\\
$w_{g}$ & \hspace{0.55cm} $\in \{0,1\}$, shutdown status of DG $g$\\
$p_{g}$ & \hspace{0.55cm} power generated above minimum \nonumber \\
& \hspace{0.55cm} by DG $g$\\
$p_g^{s}$ & \hspace{0.55cm} power from segment $s$ for DG $g$\\
$p_{n}^c$& \hspace{0.55cm} curtailed load at node $n$\\
$p^{net}_{n}$& \hspace{0.55cm} net power injection at node $n$
\end{xtabular}}}
\subsection*{\added{Parameters}}\label{3}
{
\color{change_color}{
\begin{xtabular}[h]{ll}
$\xi^{\omega}$ & \hspace{-0.25cm}net load in scenario $\omega$\\
$\pi^{\omega}$ & \hspace{-0.25cm}probability of scenario $\omega$\\
$\hat{d}_{n}$ & \hspace{-0.25cm}bid-in net demand at node $n$\\
$p_{g}^{\circ}$ & \hspace{-0.25cm}initial power generated above minimum \\
& \hspace{-0.25cm}by DG $g$\\
$u_{g}^{\circ}$ & \hspace{-0.25cm}$\in \{0,1\}$, initial commitment status of DG $g$\\
$T^{\uparrow}_{g}/T^{\downarrow}_{g}$ & \hspace{-0.25cm}minimum uptime/downtime of DG $g$\\
$T^{\uparrow, \circ}_{g}/T^{\downarrow, \circ}_{g}$ & \hspace{-0.25cm}number of hours DG $g$ has been online/offline \\
& \hspace{-0.25cm}before the scheduling horizon\\
$\overline{P}_{g}^{s}$ & \hspace{-0.25cm}maximum power available from piecewise\\
& \hspace{-0.25cm}segment $s$ for DG $g$\\
$\overline{P}_{g}/\underline{P}_{g}$ & \hspace{-0.25cm}maximum/minimum power output of\\
& \hspace{-0.25cm}DG $g$\\
$\alpha_{g}^{s}$ & \hspace{-0.25cm}cost coefficient for piecewise segment $s$ for \\
& \hspace{-0.25cm}DG $g$\\
$\alpha_{g}^{v}$ & \hspace{-0.25cm}startup cost of DG $g$\\
$\alpha_{g}^{u}$ & \hspace{-0.25cm}cost of running and operating DG $g$ \\
& \hspace{-0.25cm}at dispatch level $\underline{P}_{g}$\\
$\alpha^{c}$ & \hspace{-0.25cm}penalty cost for load curtailment\\
$\alpha^{r}$ & \hspace{-0.25cm}penalty cost for FRP shortfall\\
$\Delta_{g}^{\uparrow}/\Delta_{g}^{\downarrow}$ & \hspace{-0.25cm}ramp-up/ramp-down rate limit  \\
& \hspace{-0.25cm}of DG $g$\\
$\Delta_{g}^{\uparrow, \circ}/\Delta_{g}^{\downarrow, \circ}$ & \hspace{-0.25cm}startup/shutdown rate limit of DG $g$\\
$\overline{f}_{\ell}/\underline{f}_{\ell}$ & \hspace{-0.25cm}maximum/minimum real power flow allowed\\
 & \hspace{-0.25cm}on line $\ell$\\
$\Psi^{\ell}_{n}$ & \hspace{-0.25cm}injection shift factor of line $\ell$ with respect \\
 & \hspace{-0.25cm}to node $n$
\end{xtabular}}}

%% file: Appendices/suc.tex
\added{In this appendix, we provide the detailed mathematical formulation of the Stochastic Unit Commitment (SUC) model presented in Section~\ref{sec3a}. The first stage of the SUC problem is}
\color{change_color}{
\begin{IEEEeqnarray}{l}
\underset{\begin{subarray}{c}u_g[k], v_g[k],\\ w_g[k]\end{subarray}}{\text{minimize}} \hspace{0.5cm} \sum_{k \in \mathscr{K}} \sum_{g \in \mathscr{G}} \Big[ \alpha^{u}_{g} u_{g}[k] + \alpha^{v}_{g} v_{g}[k] \Big]  \nonumber\\
 \hspace{2.6cm}+ \sum_{\omega \in \Omega} \pi^{\omega} \mathcal{Q}({x},{\xi^{\omega}}), \label{suc_detailed_obj}\\
\text{subject to} \nonumber \\
u_{g}[k]-u_{g}[k-1] = v_{g}[k] - w_{g}[k],\nonumber \\
\hspace{2.6cm}\forall k \in \mathscr{K}\setminus\{1\}, \label{suc_detailed_st} \\
u_{g}[k]-u_{g}^{\circ} = v_{g}[k] - w_{g}[k]\nonumber \\ 
\hspace{2.6cm} \forall k \in \{1\}, \label{suc_detailed_st1}\\
\sum_{k'=k-(T^{\uparrow}_{g}\times K)+1}^{k} \hspace{-0.5cm}v_{g}[k'] \leq u_{g}[k] \nonumber\\
\hspace{2.6cm} \forall k \in \{(T^{\uparrow}_{g}\times K),\ldots,(24\times K)\},\label{suc_detailed_upt}  \\
\sum_{k'=1}^{\min\{u_g^{\circ}\times K \times(T^{\uparrow}_{g} - T^{\uparrow, \circ}_{g}), \,24 \times K\}} \hspace{-0.5cm}w_{g}[k'] = 0, \label{suc_detailed_iupt}  \\
\sum_{k'=k-(T^{\downarrow}_{g}\times K)+1}^{k} w_{g}[k'] \leq1- u_{g}[k]\nonumber\\
\hspace{2.6cm} \forall k \in \{(T^{\downarrow}_{g}\times K ),\ldots,(24\times K)\}, \label{suc_detailed_dwt}  \\
\sum_{k'=1}^{\min\{(1-u_g^{\circ})\times K \times (T^{\downarrow}_{g} - T^{\downarrow, \circ}_{g}), \,24 \times K\}} \hspace{-1.0cm}v_{g}[k'] = 0, \label{suc_detailed_idwt}  \\
u_{g}[k], v_{g}[k], w_{g}[k]  \in \{0,1\}  
\nonumber \\
\hspace{2.6cm} \forall k \in \mathscr{K} ,\label{suc_detailed_bin}
\end{IEEEeqnarray}}
\added{where \eqref{suc_detailed_st}--\eqref{suc_detailed_bin} hold for all DGs $g \in \mathscr{G}$. The objective \eqref{suc_detailed_obj} of the first stage is to minimize the commitment and startup costs plus the expected dispatch and load curtailment costs.  We enforce by (\ref{suc_detailed_st},~\ref{suc_detailed_st1}) the logical constraints that relate the variables $u[k]$, $v[k]$, and $w[k]$, (\ref{suc_detailed_upt},~\ref{suc_detailed_iupt}) the minimum uptime, and (\ref{suc_detailed_dwt},~\ref{suc_detailed_idwt}) the minimum downtime constraints.}\par
\added{For a specific vector of first-stage decision variables $x$ and a net load scenario $\xi^{\omega}$, the value function $\mathcal{Q}({x}, \xi^{\omega})$ is computed by solving the following second-stage problem:}
\color{change_color}{
\begin{IEEEeqnarray}{l}
\underset{\begin{subarray}{c}p_g[k], p_g^{s}[k], \\p_{n}^c[k]\end{subarray}}{\text{minimize}} \hspace{0.1cm} \sum_{k \in \mathscr{K}}\bigg[ \sum_{g \in \mathscr{G}} \sum_{s \in \mathscr{S}_g} {\alpha}^{s}_{g}p^{s}_{g}[k] + \sum_{n \in \mathscr{N}} \alpha^{c}p_{n}^c[k]\bigg], \label{suc_detailed_obj2}\\
\text{subject to} \nonumber \\
0 \leq p_{g}[k] \leq (\overline{P}_{g}-\underline{P}_{g})u_{g}[k]\nonumber \\ 
\hspace{3.7cm}\forall k \in \mathscr{K},\label{suc_detailed_gl} \\
p_g[k] \leq p_{g}^{\circ}+\Delta_g^{\uparrow} u_{g}^{\circ}+(\Delta_g^{\uparrow, 0} -\underline{P}_g) v_g[k],\nonumber \\
\hspace{3.7cm} \forall k \in \{1\}\label{suc_detailed_rul1} \\
p_g[k] \geq p_{g}^{\circ}-\Delta_g^{\downarrow} u_{g}^{\circ}+\Big(\Delta_g^{\downarrow}-p_{g}^{\circ}\Big) w_g[k]\nonumber \\
\hspace{3.7cm} \forall k \in \{1\}\label{suc_detailed_rll1}\\
p_g[k] \leq p_g[k-1]+\Delta_g^{\uparrow} u_g[k-1] \nonumber \\
\hspace{1.5cm}+(\Delta_g^{\uparrow, 0} -\underline{P}_g) v_g[k]\nonumber \\
\hspace{3.7cm} \forall k \in \mathscr{K} \setminus\{1\} \label{suc_detailed_rul} \\
p_g[k] \geq p_g[k-1]-\Delta_g^{\downarrow} u_g[k-1]\nonumber \\
\hspace{1.5cm}+\Big(\Delta_g^{\downarrow}-p_g[k-1]\Big) w_g[k]\nonumber \\
\hspace{3.7cm} \forall k \in \mathscr{K} \setminus\{1\} ,\label{suc_detailed_rll}\\
p_g[k] \leq w_g[k+1] (\Delta^{\downarrow, 0}_g - \underline{P}_g)\nonumber \\
\hspace{1.5cm}+\Big(1-w_g[k+1]\Big)\left(\overline{P}_g-\underline{P}_g\right)\nonumber \\
\hspace{3.7cm} \forall k \in \mathscr{K} \setminus\{K\} ,\label{suc_detailed_sd}\\
p_{g}[k] = \sum_{s \in \mathscr{S}_g} p_{g}^{s}[k]\nonumber \\
\hspace{3.7cm}\forall k \in \mathscr{K},\label{suc_detailed_ls}\\
0 \leq p^{s}_{g}[k] \leq \overline{P}^{s}_{g} - \overline{P}^{\,s-1}_{g} \nonumber \\
\hspace{3.7cm}\forall {s} \in \mathscr{S}_g,\, \forall k \in \mathscr{K},\label{suc_detailed_lsl}\\
p^{net}_{n}[k] = \sum_{g \in \mathscr{G}_{n}} p_{g}[k] + p_{n}^{c}[k] - {\xi}^{\omega}_{n}[k]\nonumber \\
\hspace{3.7cm} \forall n \in \mathscr{N}, \forall k \in \mathscr{K},\label{suc_detailed_netpower}\\
\sum_{n \in \mathscr{N}} p^{net}_{n}[k]  = 0\nonumber \\
\hspace{3.7cm} \forall k \in \mathscr{K},\label{suc_detailed_pb}\\
\underline{f}_{\ell} \leq \sum_{n \in \mathscr{N}}  \Psi^{\ell}_{n}p^{net}_{n}[k] \leq \overline{f}_{\ell}\nonumber \\
\hspace{3.7cm} \forall \ell \in \mathscr{L}, \forall k \in \mathscr{K}, \label{suc_detailed_pf}\\
p_{n}^c[k] \geq 0 \nonumber \\
\hspace{3.7cm} \forall k \in \mathscr{K}, \label{suc_detailed_nn}
\end{IEEEeqnarray}}
\added{where \eqref{suc_detailed_gl}--\eqref{suc_detailed_lsl} hold for all DGs $g \in \mathscr{G}$. The objective \eqref{suc_detailed_obj2} of the second-stage problem is to minimize the dispatch costs of DGs and the penalty cost incurred due to load curtailment. We enforce by \eqref{suc_detailed_gl}--\eqref{suc_detailed_sd} the generation and ramping limits based on the formulation laid out in \cite{morlatram}. The constraints on the power from each linear segment are stated in \eqref{suc_detailed_ls}--\eqref{suc_detailed_lsl}. We express the net real power injection at each node $n \in \mathscr{N}$ in \eqref{suc_detailed_netpower} with the convention that $p^{net}_{n}[k]>0$ if real power is injected into the system and state the system-wide power balance constraint in \eqref{suc_detailed_pb}. We use the DC power flow model to state the transmission constraints and utilize injection shift factors ({ISF}s) for network representation \cite{vanhorn}. In \eqref{suc_detailed_pf}, we express the real power flow on each line $\ell$ in terms of nodal injections and {ISF}s and constrain it to be within its line flow limits. Finally, \eqref{suc_detailed_nn} ensures that the power curtailment values $p_{n}^c[k]$ be nonnegative.}

%% file: bios.tex
\begin{IEEEbiography}[{\includegraphics[width=1in,height=1.25in,clip,keepaspectratio]{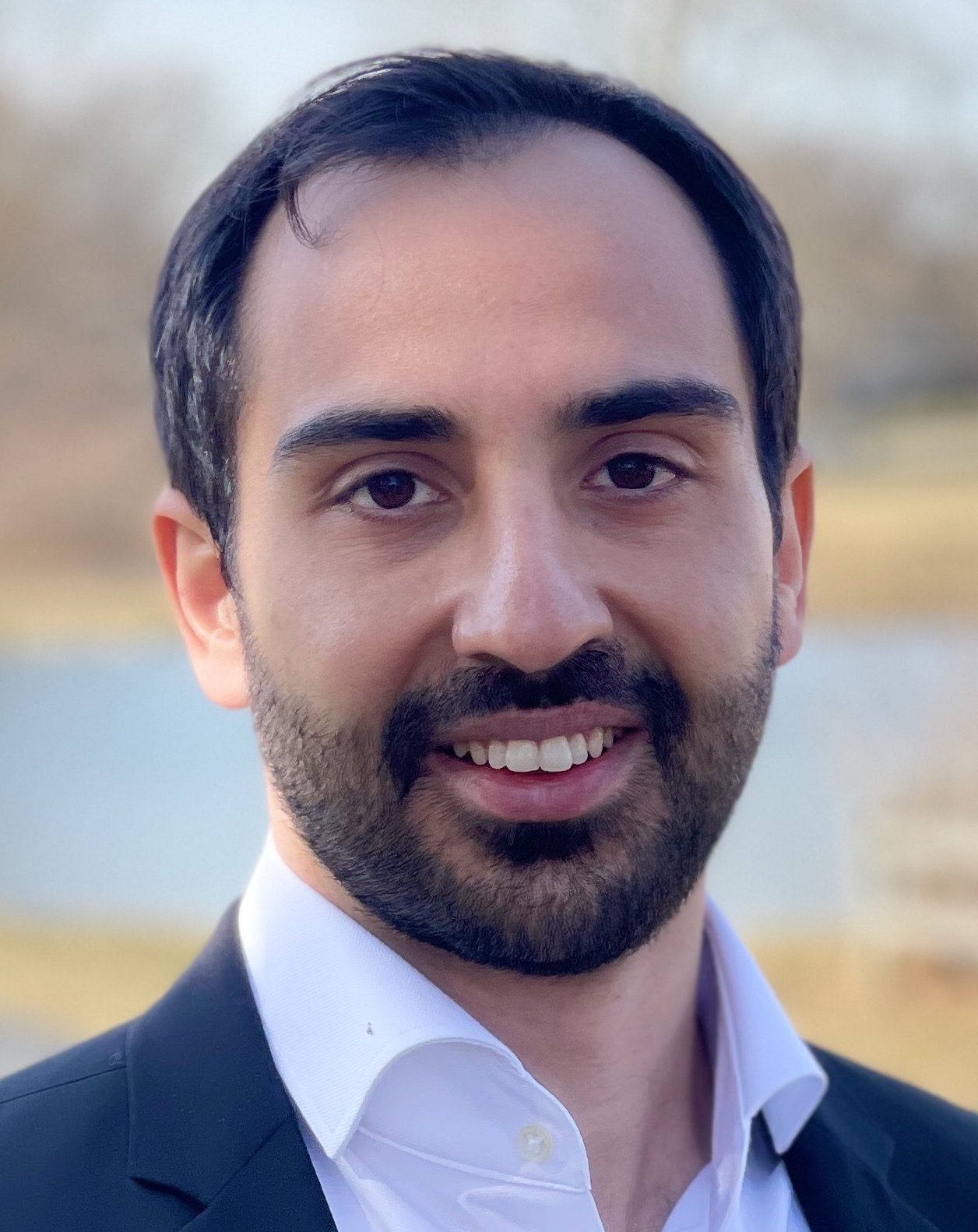}}]{Og\"un Yurdakul}
(S’20) received the B.S. degree in electrical and electronics engineering from Bogazici University, Istanbul, Turkey in 2016, the M.S. degree in electrical and computer engineering from the University of Illinois at Urbana-Champaign, IL, USA in 2018, and the Ph.D. degree in electrical engineering and computer science from the Technical University of Berlin, Berlin, Germany in 2023.\par 
Currently, he is a Quantitative Researcher at the Trailstone Group, Berlin, Germany, where he focuses on transmission trading in European and North American electricity markets and develops algorithmic trading models for European energy markets. Previously, he was with the Energy Systems and Infrastructure Analysis Division at the Argonne National Laboratory, Lemont, IL, USA. His research interests include power system operations, planning, and economics; optimization under uncertainty, and time-series forecasting for power systems applications. Dr. Yurdakul was a Fulbright Scholar during his studies at the University of Illinois at Urbana-Champaign.
\end{IEEEbiography}
\vspace{-0.5cm}
\begin{IEEEbiography}[{\includegraphics[width=1in,height=1.25in,clip,keepaspectratio]{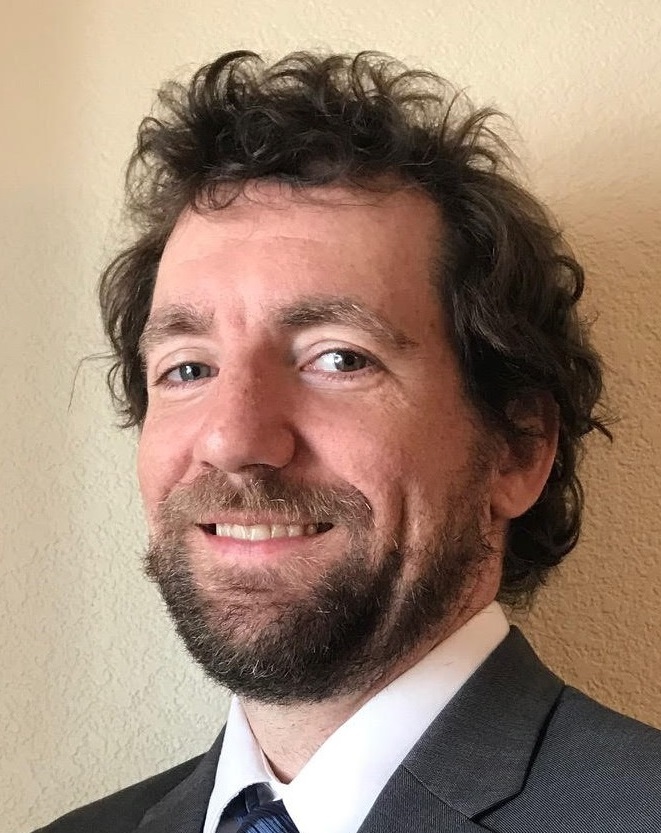}}]{Erik Ela}
is a Technical Executive and Program Manager at the Electric Power Research Institute (EPRI) working on power systems and integration for 20 years. In his role, he provides technical leadership in several areas around the evolution of power systems and electricity markets. He facilitates the R\&D collaborative group of technical experts of North America’s Independent System Operators (ISOs) and Regional Transmission Organizations (RTOs) EPRI’s electricity market design and operations research program. He has worked with the NREL and NYISO, and received his BS, MS, and PhD degrees in Electrical Engineering.
\end{IEEEbiography}
\vspace{-0.5cm}
\begin{IEEEbiography}[{\includegraphics[width=1in,height=1.25in,clip,keepaspectratio]{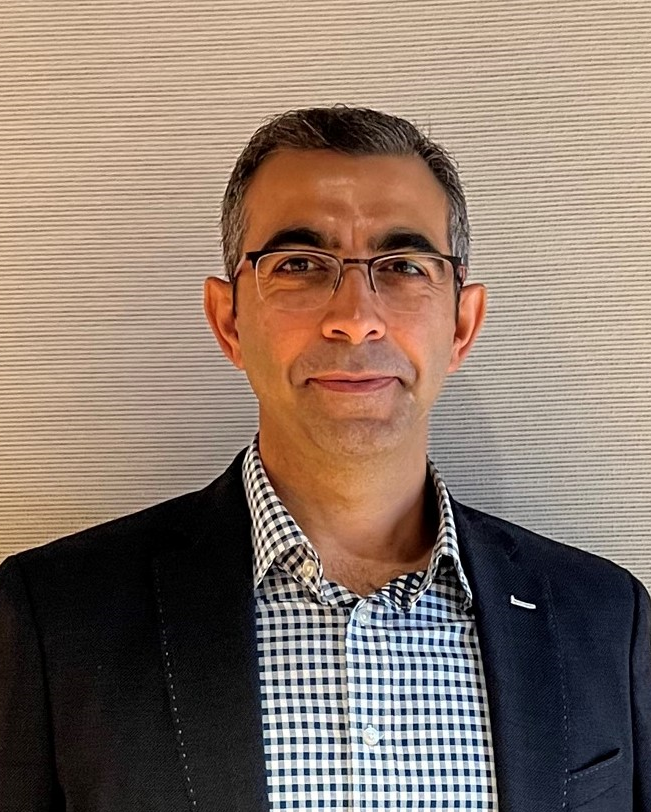}}]{Farhad Billimoria}
is a Visiting Research Fellow at the Oxford Institute for Energy Studies, and a Director in the power consulting team at S\&P Global. His research focuses upon market design for low-carbon power systems, including the flexibility markets, resource adequacy and reliability mechanisms, hedging and financial products, and markets for power system security. He has a PhD from the University of Oxford, and a Masters of Energy Systems from the University of Melbourne.
\end{IEEEbiography}